%% file: main.tex
\documentclass[a4paper,UKenglish,cleveref,autoref,thm-restate]{lipics-v2021}

\usepackage[utf8]{inputenc}
\usepackage[table]{xcolor}
\usepackage{amssymb,amsmath}
\usepackage{url}
\usepackage{float}
\usepackage{complexity}
\usepackage{hyperref}
\usepackage[ruled,vlined,linesnumbered,noend]{algorithm2e}
\usepackage{tikz}
\usetikzlibrary{calc,arrows.meta,decorations.pathreplacing}
\input{figures/tikzdefs}
\usepackage{framed}
\usepackage{xspace}
\usepackage{xifthen}
\usepackage{tabularx}
\usepackage{booktabs}
\usepackage{array}
\usepackage{makecell}

\hypersetup{colorlinks, linkcolor={dark-blue}, citecolor={mid-green}, urlcolor={dark-blue}}
\definecolor{mid-green}{rgb}{0.15,0.65,0.15}
\definecolor{dark-blue}{rgb}{0.15,0.15,0.9}
\definecolor{dark-red}{rgb}{0.7,0.15,0.15}

\newtheorem{construction}{Construction}

\newlength{\RoundedBoxWidth}
\newsavebox{\GrayRoundedBox}
\newenvironment{GrayBox}[1]%
   {\setlength{\RoundedBoxWidth}{.93\textwidth}
    \def\boxheading{#1}
    \begin{lrbox}{\GrayRoundedBox}
       \begin{minipage}{\RoundedBoxWidth}}%
   {   \end{minipage}
    \end{lrbox}
    \begin{center}
    \begin{tikzpicture}%
       \node(Text)[draw=black!20,fill=white,rounded corners,%
             inner sep=2ex,text width=\RoundedBoxWidth]%
             {\usebox{\GrayRoundedBox}};
        \coordinate(x) at (current bounding box.north west);
        \node [draw=white,rectangle,inner sep=3pt,anchor=north west,fill=white]
        at ($(x)+(6pt,.75em)$) {\boxheading};
    \end{tikzpicture}
    \end{center}}
\newenvironment{defproblemx}[2][]{\noindent\ignorespaces%
                \FrameSep=6pt\parindent=0pt\vspace*{-1.5em}%
                \begin{GrayBox}{\textsc{#2}}%
                \begin{tabular*}{\textwidth}{@{\hspace{.1em}} >{\itshape} p{1.8cm} p{0.8\textwidth} @{}}%
            }{\end{tabular*}\end{GrayBox}\ignorespacesafterend}
\newcommand{\defproblema}[3]{%
  \begin{defproblemx}{#1}
    {\bf Instance:}  & #2 \\
    {\bf Question:} & #3
  \end{defproblemx}}

\renewcommand{\NP}{{\sf NP}\xspace}
\renewcommand{\P}{{\sf P}\xspace}
\renewcommand{\APX}{{\sf APX}\xspace}
\newcommand{\dd}{d}                         
\newcommand{\ellp}{\ell}                    
\newcommand{\lea}{\preceq}                  
\newcommand{\LB}{{\sf LB}}
\newcommand{\med}{{\sf med}}
\newcommand{\clo}{{\sf clo}}
\newcommand{\Subd}{{\rm Subd}_2}

\title{The {\sc Swap Median} and {\sc Swap Closest} Problems are \NP-complete for Three Permutations}

\author{Lu\'is Cunha}
{Instituto de Computação, Universidade Federal Fluminense, Brasil \and \url{http://www.ic.uff.br/~lfignacio}}
{lfignacio@ic.uff.br}
{https://orcid.org/0000-0002-3797-6053}
{FAPERJ-JCNE (E-26/201.372/2022), and CNPq-Universal (406173/2021-4).}

\author{Thiago Nascimento}
{Instituto de Computação, Universidade Federal Fluminense, Brasil}
{thiago\_nascimento@id.uff.br}
{https://orcid.org/0009-0005-5278-6679}
{}

\author{Arnaud Mary}
{Laboratoire de Biométrie et Biologie Évolutive, Université Claude Bernard Lyon 1, France}
{arnaud.mary@univ-lyon1.fr}
{}
{}

\authorrunning{L. Cunha, T. Nascimento and A. Mary}
\titlerunning{Swap Median and Swap Closest are NP-complete for Three Permutations}
\Copyright{Luís Cunha, Thiago Nascimento, and Arnaud Mary}
\ccsdesc[500]{Theory of computation~Problems, reductions and completeness}
\ccsdesc[300]{Applied computing~Computational genomics}
\keywords{Swap distance, Cayley distance, median permutation, closest permutation, genome rearrangements, absolute order, noncrossing partitions, NP-completeness, APX-hardness}
\relatedversion{}
\EventEditors{}
\EventLongTitle{}
\EventShortTitle{}
\EventAcronym{}
\EventYear{}
\EventDate{}
\EventLocation{}
\EventLogo{}
\SeriesVolume{}
\ArticleNo{}

\begin{document}
\nolinenumbers
\maketitle

\input{sections/abstract}
\input{sections/introduction}
\input{sections/preliminaries}
\input{sections/geodesic}
\input{sections/pivot_hardness}
\input{sections/closest}
\input{sections/median}
\input{sections/algorithms}
\input{sections/approximation}
\input{sections/experiments}
\input{sections/conclusion}

\bibliography{references}

\appendix
\input{sections/appendix}

\end{document}

%% file: figures/tikzdefs.tex
\makeatletter
\newcommand{\drawcycle}[6]{%
  \draw[#6, line width=0.9pt] (#2,#3) circle (#4);
  \foreach \lab [count=\n] in {#5} {\xdef\cyclen{\n}}%
  \foreach \lab [count=\i] in {#5} {%
    \pgfmathsetmacro{\ang}{90-360*(\i-1)/\cyclen}%
    \coordinate (#1-\lab) at ($(#2,#3)+(\ang:#4)$);
    \fill[#6] (#1-\lab) circle (1.3pt);
    \node[font=\scriptsize, inner sep=1pt] at ($(#2,#3)+(\ang:#4+0.22)$) {$\lab$};
  }%
}
\makeatother
\tikzset{
  chord/.style={line width=1.1pt, black},
  chordsel/.style={line width=1.8pt, black!60!green},
  chordoff/.style={line width=0.8pt, black, dashed},
  gvert/.style={circle, draw, fill=white, inner sep=1.2pt, font=\scriptsize, minimum size=12pt},
  gsel/.style={circle, draw=black!60!green, fill=green!15, line width=1pt, inner sep=1.2pt, font=\scriptsize, minimum size=12pt},
  redc/.style={red!75!black},
  bluec/.style={blue!70!black},
}

%% file: sections/abstract.tex
\begin{abstract}
The swap distance (or Cayley distance) between two permutations is the minimum number of exchanges of two elements transforming one into the other. Given three permutations, {\sc Swap Median} asks for a permutation minimizing the sum of the swap distances to the inputs, and {\sc Swap Closest} asks for one minimizing the maximum distance. Both problems are \NP-hard for an unbounded number of inputs, but the case of three inputs, the smallest non-trivial one, has remained open. We prove that both problems are \NP-complete for three permutations. {\sc Swap Median} remains \NP-complete even when two of the inputs differ by disjoint swaps. For {\sc Swap Closest}, it is already \NP-complete to decide whether the trivial lower bound is attained. Along the way, we show that finding a permutation lying on shortest paths from one input to each of the other two, as far as possible from that input, is \NP-hard and \APX-hard.

We also characterize when the triangle-inequality lower bound of the median is attained, and we give a quadratic-time algorithm when cycles are short, which yields a dichotomy. Finally, experiments on simulated evolutionary data show that a simple certified heuristic is optimal on all small instances and recovers ancestral permutations.
\end{abstract}

%% file: sections/introduction.tex
\section{Introduction}

Genome rearrangement problems study large-scale mutations that modify the order of conserved genomic markers. They provide a classical combinatorial model in comparative genomics and have been extensively investigated in computational biology and theoretical computer science~\cite{fertin2009combinatorics,pevzner2000computational,watterson1982chromosome}. In the simplest setting, each genome is represented by a permutation: all genomes contain the same set of genes, each gene occurs exactly once, and the order of the genes in a single chromosome is known.

A genome rearrangement distance is defined by choosing a set of allowed operations and asking for the minimum number of such operations needed to transform one permutation into another. In this paper we focus on the swap distance. A swap exchanges the contents of two positions of a permutation written in one-line notation; equivalently, it composes the permutation with a transposition. The swap distance between two permutations is the minimum number of swaps required to transform one into the other; it equals the size of the permutations minus the number of cycles of the relative permutation, and is therefore computable in linear time~\cite{eriksen2007reversal,fertin2009combinatorics}. The same metric is known in statistics and machine learning as the \emph{Cayley distance}~\cite{diaconis1988group}, where it is one of the standard distances underlying distance-based ranking models such as the Mallows model~\cite{fligner1986distance,irurozki2018cayley}. In that context, computing a consensus (central) permutation of a sample under the Cayley distance is exactly the swap median problem studied here.

When more than two genomes are considered, pairwise distances are often not sufficient. Two central consensus problems are the {\sc Median} and {\sc Closest} problems. Given a set of input permutations and a distance, the {\sc Median} problem asks for a permutation minimizing the sum of the distances to the inputs, and the {\sc Closest} problem asks for a permutation minimizing the maximum distance to the inputs. These problems are important in ancestral reconstruction and in the design of consensus genomic representations~\cite{bader2011transposition,caprara2003reversal,cunha2020computational,cunha2024parameterized,haghighi2012medians,pe1998median,tannier2009multichromosomal}.

The computational complexity of median and closest problems depends strongly on the underlying distance. The median problem under the single-cut-or-join distance can be solved in polynomial time~\cite{feijao2011scj}, whereas the breakpoint median~\cite{bryant1998complexity,pe1998median}, the reversal median~\cite{caprara2003reversal}, the transposition median~\cite{bader2011transposition} and the block-interchange median~\cite{cunha2023complexity} are \NP-hard already for three input permutations. For the swap distance, Popov~\cite{popov2007multiple} proved \NP-hardness of multiple genome rearrangement by swaps when the number of input permutations is part of the input, and Bachmaier et al.~\cite{bachmaier2015hardness} obtained \NP-completeness of the closest problem (maximum rank aggregation) for an unbounded number of permutations under several metrics, including the Cayley distance. Eriksen~\cite{eriksen2007reversal} studied structural properties of swap medians, and the complexity of {\sc Swap Median} with three inputs is explicitly listed as open in~\cite{cunha2023complexity}. Three is the smallest non-trivial number of inputs, and we settle this case for both problems: with two inputs any permutation on a shortest path between them is an optimal median, and the midpoint of such a path is an optimal closest permutation. For Kendall's tau distance, the analogous question (Kemeny rank aggregation with three voters) has just been settled: Madarasi~\cite{madarasi2026kemeny} and Peters~\cite{peters2026kemeny} independently proved that it is \NP-hard.

\paragraph*{Our contributions.}
We study {\sc Swap Median} and {\sc Swap Closest} with three input permutations through the geometry of shortest paths (geodesics). A permutation lies on a geodesic from the identity to $\tau$ if and only if it is below $\tau$ in the \emph{absolute order} of the symmetric group, and the intervals of this order are products of lattices of noncrossing partitions~\cite{biane1997crossings,brady2001partial}. Choosing an input $\pi_r$ as a \emph{pivot}, the permutations lying simultaneously on a geodesic from $\pi_r$ to each of the two other inputs correspond to common lower bounds of two permutations in the absolute order. Our results are the following.

\begin{itemize}
    \item \emph{Geodesic structure (\autoref{sec:geodesic}).} The triangle-inequality lower bound $\LB$ for the median is attained if and only if the maximum rank $M_r$ of a common lower bound equals the Gromov product $\beta_r$ of the inputs at the pivot, for some (equivalently, every) pivot $r$. Every permutation whose median value is at most $\LB+1$ is a pivot solution. In general, however, optimal medians are not pivot solutions, and we give explicit counterexamples, including a connected one. We also give a crossing graph whose independent sets yield common lower bounds; pairs of swaps sharing an element must be treated as conflicting.

    \item \emph{Hardness of the pivot problem (\autoref{sec:pivot}).} Deciding whether $M_1\geq k$, that is, whether there is a permutation at distance at least $k$ from $\pi_1$ lying on a geodesic from $\pi_1$ to $\pi_2$ and on a geodesic from $\pi_1$ to $\pi_3$, is \NP-complete. The maximization version is \APX-hard. This holds even when all cycles of the two relative permutations have length at most $8$ and at most $4$, respectively.

    \item \emph{{\sc Swap Closest} (\autoref{sec:closest}).} {\sc Swap Closest} is \NP-complete for three input permutations. It is even \NP-complete to decide whether the closest lower bound, half of the largest pairwise distance, is attained. This strengthens the \NP-hardness for an unbounded number of inputs~\cite{bachmaier2015hardness,popov2007multiple}. The proof uses padding blocks that make two triangle inequalities tight.

    \item \emph{{\sc Swap Median} (\autoref{sec:median}).} {\sc Swap Median} is \NP-complete for three input permutations, even when two of them differ by a set of disjoint swaps. The proof has three steps. First, a projection lemma: when $\pi_1^{-1}\pi_2$ is a product of disjoint transpositions, some optimal median lies in the Boolean cube $I(\pi_1,\pi_2)$; the proof uses the Cohn--Lempel formula~\cite{cohn1972cycle}. Second, optimizing over this cube is the same as maximizing the number of closed trails in a transition system of an Eulerian digraph. Third, this maximization is \NP-hard by Colbourn's theorem on partial Latin squares~\cite{colbourn1984complexity}.

    \item \emph{Polynomial cases (\autoref{sec:algorithms}).} If one of the two relative permutations has all cycles of length at most three, $M_r$ can be computed in $O(n^2)$ time by dynamic programming over noncrossing structures. Combined with the hardness result, this gives a dichotomy with respect to cycle lengths. As a consequence, attainment of the median lower bound can be decided in polynomial time on such instances.

    \item \emph{Approximation (\autoref{sec:approximation}).} Any common lower bound computed at a pivot yields an \emph{a posteriori} certificate of the form $\textrm{value}\leq \LB+(\beta_r-\ellp(\lambda))$. We formulate a robustness conjecture. If it holds, {\sc Swap Closest} with three inputs is \APX-hard, and exact pivot optimization approximates {\sc Swap Median} within $10/9$.

    \item \emph{Experiments (\autoref{sec:experiments}).} On simulated data, including evolutionary scenarios, we compare the pivot heuristic and an iterated local search with exact optima (obtained by exhaustive search for $n\leq 10$) and with lower bounds for larger $n$. The implementation is publicly available.
\end{itemize}

The paper is organized as follows. \autoref{sec:preliminares} introduces notation, the absolute order and the problems. \autoref{sec:geodesic} develops the geodesic structure of three permutations. \autoref{sec:pivot} proves the hardness of the pivot problem, \autoref{sec:closest} proves the \NP-completeness of {\sc Swap Closest}, and \autoref{sec:median} proves the \NP-completeness of {\sc Swap Median}. \autoref{sec:algorithms} gives polynomial-time algorithms, \autoref{sec:approximation} discusses approximation, and \autoref{sec:experiments} reports experiments. \autoref{sec:conclusion} lists open problems.

%% file: sections/preliminaries.tex
\section{Preliminaries}
\label{sec:preliminares}

For a positive integer $n$, let $[n]=\{1,2,\ldots,n\}$. A \emph{permutation} of size $n$ is a bijection from $[n]$ to itself, written in one-line notation as $\pi=[\pi(1)\ \pi(2)\ \cdots\ \pi(n)]$. The identity is $\iota=[1\ 2\ \cdots\ n]$. Permutations are composed from right to left, $(\pi\sigma)(i)=\pi(\sigma(i))$.

\paragraph*{Cycles and swaps.}
The \emph{cycle} of $\pi$ containing $x$ is $(x\ \pi(x)\ \pi^2(x)\ \cdots\ \pi^{k-1}(x))$, where $k$ is the smallest positive integer with $\pi^k(x)=x$. Every permutation decomposes uniquely into disjoint cycles; we denote by $c(\pi)$ the number of cycles, fixed points included. For example, $\pi=[8\ 5\ 1\ 3\ 2\ 7\ 6\ 4]$ has cycle decomposition $(1\ 8\ 4\ 3)(2\ 5)(6\ 7)$, so $c(\pi)=3$ (\autoref{fig:alggraph}).

\begin{figure}[!h]
    \centering
    \input{figures/fig_cyclegraph}
    \caption{(a)~The permutation $\pi=[8\ 5\ 1\ 3\ 2\ 7\ 6\ 4]$ in two-line notation. (b)~Its cycle graph, with an arc $x\to\pi(x)$ for every $x$; each cycle is drawn clockwise on a circle, as in all later figures. (c)~Exchanging the values $1$ and $4$ (the boxed entries, at positions $3$ and $8$) gives $(1\ 4)\pi=\pi(3\ 8)$. Since $1$ and $4$ lie in the same cycle, the chord $\{1,4\}$ (dashed) splits it into $(1\ 8)$ and $(4\ 3)$ (\autoref{lem:swap-effect}).}
    \label{fig:alggraph}
\end{figure}

For distinct $a,b\in[n]$, let $(a\ b)$ denote the transposition exchanging $a$ and $b$. A \emph{swap} applied to $\pi$ exchanges the entries at two positions $i,j$ of its one-line notation; the result is $\pi(i\ j)$. Equivalently, exchanging the \emph{values} $a=\pi(i)$ and $b=\pi(j)$ gives $(a\ b)\pi=\pi(i\ j)$. The \emph{swap distance} $\dd(\pi,\sigma)$ is the minimum number of swaps transforming $\pi$ into $\sigma$, and we write $\ellp(\pi)=\dd(\iota,\pi)$. The following facts are classical (see, e.g.,~\cite{diaconis1988group,fertin2009combinatorics}).

\begin{lemma}\label{lem:swap-effect}
Let $\alpha$ be a permutation and $a\neq b$. If $a$ and $b$ lie in a common cycle $(a\ x_1\ \cdots\ x_p\ b\ y_1\ \cdots\ y_q)$ of $\alpha$, then $(a\ b)\alpha$ is obtained from $\alpha$ by replacing this cycle with the two cycles $(a\ x_1\ \cdots\ x_p)$ and $(b\ y_1\ \cdots\ y_q)$; we say that the swap \emph{splits} the cycle. Otherwise, $(a\ b)\alpha$ is obtained by merging the two cycles containing $a$ and $b$. In particular, $c((a\ b)\alpha)=c(\alpha)\pm 1$. Consequently, $\ellp(\pi)=n-c(\pi)$ and
\[
\dd(\pi,\sigma)=\ellp(\pi^{-1}\sigma)=n-c(\pi^{-1}\sigma).
\]
\end{lemma}

Note that the two cycles produced by a split are arcs of the original cycle, each with the cyclic order induced by it. Since $\pi^{-1}\sigma$ and $\sigma\pi^{-1}$ are conjugate, $c(\pi^{-1}\sigma)=c(\sigma\pi^{-1})$, so the formula agrees with the usual definition through the relative permutation $\sigma\pi^{-1}$. The swap distance is a metric on the symmetric group $S_n$; it is \emph{left-invariant}, $\dd(\lambda\pi,\lambda\sigma)=\dd(\pi,\sigma)$, which allows us to assume that one input permutation is the identity. Moreover, $\dd(\pi,\sigma)$ has the parity of the permutation $\pi^{-1}\sigma$.

\paragraph*{Geodesics and the absolute order.}
A permutation $\sigma$ lies on a \emph{geodesic} (a shortest path in the Cayley graph generated by transpositions) between $\pi$ and $\tau$ if $\dd(\pi,\sigma)+\dd(\sigma,\tau)=\dd(\pi,\tau)$. We denote by $I(\pi,\tau)$ the set of such permutations. The \emph{absolute order} on $S_n$ is defined by
\[
\sigma\lea\tau \iff \ellp(\sigma)+\ellp(\sigma^{-1}\tau)=\ellp(\tau)\iff \sigma\in I(\iota,\tau).
\]
By left-invariance, $\sigma\in I(\pi,\tau)$ if and only if $\pi^{-1}\sigma\lea\pi^{-1}\tau$. The absolute order is a graded partial order with rank function $\ellp$. Its lower intervals are described by noncrossing partitions. Let $C=(c_1\ c_2\ \cdots\ c_k)$ be a cycle. Two disjoint subsets $X,Y$ of $\{c_1,\ldots,c_k\}$ \emph{cross} in $C$ if there are $a,b\in X$ and $x,y\in Y$ appearing in the cyclic order $a,x,b,y$ along $C$; a partition of $\{c_1,\ldots,c_k\}$ is \emph{noncrossing} with respect to $C$ if no two of its blocks cross.

\begin{theorem}[Biane~\cite{biane1997crossings}, Brady~\cite{brady2001partial}]\label{thm:biane}
Let $\sigma,\tau\in S_n$. Then $\sigma\lea\tau$ if and only if
(i) every cycle of $\sigma$ is contained in a cycle of $\tau$;
(ii) for every cycle $C$ of $\tau$, the element sets of the cycles of $\sigma$ contained in $C$ form a noncrossing partition with respect to $C$; and
(iii) every cycle of $\sigma$ lists its elements in the cyclic order induced by the cycle of $\tau$ containing it.
Consequently, the interval $[\iota,\tau]$ is isomorphic to the product, over the cycles $C$ of $\tau$, of the lattices of noncrossing partitions of $|C|$ elements.
\end{theorem}

We will use three simple consequences. First, intervals are graded: if $\sigma\lea\tau$ and $0\leq k\leq \ellp(\sigma)$, then there is $\sigma'\lea\sigma$ (hence $\sigma'\lea\tau$) with $\ellp(\sigma')=k$. Second, if the elements are partitioned into blocks and $\tau$ maps every block onto itself, then every $\sigma\lea\tau$ maps every block onto itself, and $\sigma\lea\tau$ holds if and only if it holds block by block. Third, if $a,b$ lie in a common cycle $C$ of $\tau$, then $(a\ b)\lea\tau$.

\paragraph*{Median and closest problems.}

\defproblema
{Swap Median}
{Three permutations $\pi_1,\pi_2,\pi_3$ of the same size and a non-negative integer $f$.}
{Is there a permutation $\sigma$ such that $\sum_{i=1}^3 \dd(\sigma,\pi_i)\leq f$?}

\defproblema
{Swap Closest}
{Three permutations $\pi_1,\pi_2,\pi_3$ of the same size and a non-negative integer $f$.}
{Is there a permutation $\sigma$ such that $\max_{i\in\{1,2,3\}} \dd(\sigma,\pi_i)\leq f$?}

Both problems belong to \NP, since distances are computable in linear time. For an instance $\Pi=(\pi_1,\pi_2,\pi_3)$ we write $\dd_{ij}=\dd(\pi_i,\pi_j)$, $F_\Pi(\sigma)=\sum_i\dd(\sigma,\pi_i)$, and denote by $\med(\Pi)$ and $\clo(\Pi)$ the optimum values. The triangle inequality gives the lower bounds
\[
\med(\Pi)\ \geq\ \LB(\Pi)=\frac{\dd_{12}+\dd_{13}+\dd_{23}}{2},
\qquad
\clo(\Pi)\ \geq\ \LB_{\clo}(\Pi)=\left\lceil \frac{Q}{2}\right\rceil,\quad Q=\max_{i<j}\dd_{ij}.
\]
$\LB(\Pi)$ is an integer: $\pi_2^{-1}\pi_3=(\pi_1^{-1}\pi_2)^{-1}(\pi_1^{-1}\pi_3)$, so the three relative permutations cannot all be odd and $\dd_{12}+\dd_{13}+\dd_{23}$ is even. A solution \emph{attains} a lower bound if its value equals it. The term \emph{median} refers to the classical notion of a median in a metric space (a point minimizing the sum of distances), which for three points generalizes the median of three numbers.

\begin{observation}\label{obs:lower-bound-paths}
A permutation $\sigma$ satisfies $F_\Pi(\sigma)=\LB(\Pi)$ if and only if $\sigma\in I(\pi_i,\pi_j)$ for every pair $i<j$.
\end{observation}

\begin{proof}
Summing the three triangle inequalities $\dd_{ij}\leq \dd(\pi_i,\sigma)+\dd(\sigma,\pi_j)$ gives $2\LB(\Pi)\leq 2F_\Pi(\sigma)$, with equality if and only if each inequality is tight.
\end{proof}

Finally, the \emph{disjoint union} of permutations $\pi$ of size $p$ and $\sigma$ of size $q$ is $\pi\uplus\sigma=[\pi(1),\ldots,\pi(p),\sigma(1)+p,\ldots,\sigma(q)+p]$. Cycles of $\pi\uplus\sigma$ are those of $\pi$ and the shifted cycles of $\sigma$; hence $\dd(\pi_1\uplus\sigma_1,\pi_2\uplus\sigma_2)=\dd(\pi_1,\pi_2)+\dd(\sigma_1,\sigma_2)$. For $t\geq 1$, $\rho_t=[2\ 3\ \cdots\ t\ 1]$ is a single $t$-cycle and $\dd(\iota_t,\rho_t)=t-1$.

%% file: figures/fig_cyclegraph.tex
\begin{tikzpicture}[
    vx/.style={circle, draw, fill=white, inner sep=0pt, minimum size=13pt, font=\scriptsize},
    arc/.style={-{Stealth[length=4pt]}, line width=0.8pt}]
  \newcommand{\dcyc}[5]{%
    \foreach \lab [count=\n] in {#4} {\xdef\cyclen{\n}}%
    \pgfmathsetmacro{\gap}{\cyclen==1 ? 0 : (\cyclen==2 ? 34 : 22)}%
    \foreach \lab [count=\i] in {#4} {%
      \pgfmathsetmacro{\ang}{90-360*(\i-1)/\cyclen}%
      \pgfmathsetmacro{\nxt}{\ang-360/\cyclen}%
      \draw[arc, #5] ($(#1,#2)+(\ang-\gap:#3)$) arc (\ang-\gap:\nxt+\gap:#3);
      \node[vx, draw=#5] at ($(#1,#2)+(\ang:#3)$) {$\lab$};
    }%
  }
  \node[font=\small] at (-0.1,1.05) {(a)};
  \foreach \v [count=\i] in {8,5,1,3,2,7,6,4} {
    \node[font=\scriptsize, text=black!55] at (0.35*\i+0.2,0.55) {$\i$};
    \node[font=\small] at (0.35*\i+0.2,0.05) {$\v$};
  }
  \draw[black!40] (0.38,0.32) -- (3.0,0.32);
  \node[font=\scriptsize, anchor=east] at (0.42,0.55) {$x$};
  \node[font=\scriptsize, anchor=east] at (0.42,0.05) {$\pi(x)$};
  \draw[black!60!green, line width=0.9pt, rounded corners=2pt] (1.12,-0.17) rectangle (1.38,0.25);
  \draw[black!60!green, line width=0.9pt, rounded corners=2pt] (2.87,-0.17) rectangle (3.13,0.25);
  \node[font=\scriptsize, align=center] at (1.6,-0.65) {$\pi=[8\ 5\ 1\ 3\ 2\ 7\ 6\ 4]$};
  \node[font=\small] at (3.55,1.05) {(b)};
  \dcyc{4.6}{0.1}{0.62}{1,8,4,3}{black}
  \draw[black!60!green, line width=1.3pt, dashed] (4.6,0.5) -- (4.6,-0.3);
  \dcyc{6.1}{0.1}{0.42}{2,5}{black}
  \dcyc{7.1}{0.1}{0.42}{6,7}{black}
  \node[font=\scriptsize] at (5.85,-0.95) {$\pi=(1\ 8\ 4\ 3)(2\ 5)(6\ 7)$,\ \ $c(\pi)=3$};
  \draw[-{Stealth[length=5pt]}, line width=0.9pt] (7.75,0.1) -- (8.55,0.1)
     node[midway, above, font=\scriptsize] {$(1\ 4)\,\pi$};
  \node[font=\small] at (8.7,1.05) {(c)};
  \dcyc{9.4}{0.1}{0.42}{1,8}{black!60!green}
  \dcyc{10.4}{0.1}{0.42}{4,3}{black!60!green}
  \dcyc{11.4}{0.1}{0.42}{2,5}{black}
  \dcyc{12.4}{0.1}{0.42}{6,7}{black}
  \node[font=\scriptsize] at (10.9,-0.95) {$(1\ 4)\pi=[8\ 5\ 4\ 3\ 2\ 7\ 6\ 1]$,\ \ $c=4$};
\end{tikzpicture}

%% file: sections/geodesic.tex
\section{Geodesic structure of three permutations}
\label{sec:geodesic}

Throughout this section, $\Pi=(\pi_1,\pi_2,\pi_3)$ is a fixed instance. For a \emph{pivot} $r\in\{1,2,3\}$, let $\{s,t\}=\{1,2,3\}\setminus\{r\}$ and define the relative permutations $\tau^{r}_s=\pi_r^{-1}\pi_s$ and $\tau^{r}_t=\pi_r^{-1}\pi_t$, so that $\ellp(\tau^r_s)=\dd_{rs}$ and $\ellp(\tau^r_t)=\dd_{rt}$. We set
\[
D_r=\dd_{rs}+\dd_{rt},\qquad
\beta_r=\frac{\dd_{rs}+\dd_{rt}-\dd_{st}}{2}.
\]
$D_r$ is the median value of the input $\pi_r$ itself. The quantity $\beta_r$ is an integer by the parity argument of \autoref{sec:preliminares}; it is the \emph{Gromov product} of $\pi_s$ and $\pi_t$ seen from $\pi_r$, and it satisfies $D_r-\beta_r=\LB(\Pi)$ for every $r$.

\subsection{Pivot solutions and the median lower bound}

\begin{definition}\label{def:pivot}
A \emph{common lower bound} at pivot $r$ is a permutation $\lambda$ with $\lambda\lea\tau^r_s$ and $\lambda\lea\tau^r_t$. A \emph{pivot solution} at $r$ is a permutation $\sigma=\pi_r\lambda$ where $\lambda$ is a common lower bound at $r$; equivalently, $\sigma\in I(\pi_r,\pi_s)\cap I(\pi_r,\pi_t)$. We denote by $M_r$ the maximum of $\ellp(\lambda)$ over common lower bounds at $r$, by $P_r=D_r-M_r$ the \emph{pivot value}, and by $g_r=\beta_r-M_r$ the \emph{pivot gap}.
\end{definition}

In words, a pivot solution is obtained by walking from $\pi_r$ along swaps that bring us closer to both $\pi_s$ and $\pi_t$, and $M_r$ is the maximum length of such a walk.

\begin{lemma}\label{lem:pivot-value}
If $\lambda$ is a common lower bound at pivot $r$, then $F_\Pi(\pi_r\lambda)=D_r-\ellp(\lambda)$.
\end{lemma}

\begin{proof}
By left-invariance, $\dd(\pi_r,\pi_r\lambda)=\ellp(\lambda)$ and $\dd(\pi_r\lambda,\pi_s)=\ellp(\lambda^{-1}\tau^r_s)=\ellp(\tau^r_s)-\ellp(\lambda)=\dd_{rs}-\ellp(\lambda)$, and similarly for $\pi_t$. Summing gives $\ellp(\lambda)+\dd_{rs}+\dd_{rt}-2\ellp(\lambda)$.
\end{proof}

\begin{theorem}\label{thm:lb-characterization}
For every instance $\Pi$ and every pivot $r$:
\begin{enumerate}
\item $M_r\leq \beta_r$, that is, $g_r\geq 0$;
\item $\med(\Pi)=\LB(\Pi)$ if and only if $M_r=\beta_r$; in particular, $g_1=0 \iff g_2=0\iff g_3=0$;
\item $\LB(\Pi)\leq \med(\Pi)\leq \min_r P_r=\LB(\Pi)+\min_r g_r\leq \min_r D_r\leq \frac43\LB(\Pi)$.
\end{enumerate}
\end{theorem}

\begin{proof}
Let $\lambda$ be a common lower bound with $\ellp(\lambda)=M_r$. By \autoref{lem:pivot-value}, $\LB\leq F_\Pi(\pi_r\lambda)=D_r-M_r$, so $M_r\leq D_r-\LB=\beta_r$, proving (1). If $M_r=\beta_r$, then $F_\Pi(\pi_r\lambda)=D_r-\beta_r=\LB$. Conversely, if $\sigma$ attains $\LB$, then $\sigma\in I(\pi_r,\pi_s)\cap I(\pi_r,\pi_t)$ by \autoref{obs:lower-bound-paths}, so $\lambda=\pi_r^{-1}\sigma$ is a common lower bound and $\LB=F_\Pi(\sigma)=D_r-\ellp(\lambda)$, i.e., $\ellp(\lambda)=\beta_r$; by (1), $M_r=\beta_r$. This proves (2), since the condition $\med(\Pi)=\LB(\Pi)$ does not depend on $r$. For (3), the first inequality is the triangle inequality, the second follows from \autoref{lem:pivot-value}, the identity follows from $D_r-\beta_r=\LB$, the fourth inequality uses $M_r\geq 0$, and the last one follows from $D_1+D_2+D_3=2(\dd_{12}+\dd_{13}+\dd_{23})=4\LB$.
\end{proof}

The bound $M_r\leq\beta_r$ is the reason why $\beta_r$ cannot be chosen freely in a reduction: it is an upper bound that holds for every instance. The next result shows that solutions which are almost optimal with respect to the lower bound are always pivot solutions.

For a permutation $\sigma$ and $i<j$, let $e_{ij}(\sigma)=\dd(\pi_i,\sigma)+\dd(\sigma,\pi_j)-\dd_{ij}\geq 0$ be the \emph{defect} of $\sigma$ with respect to the pair $\{i,j\}$. Summing the three defects gives
\begin{equation}\label{eq:defects}
F_\Pi(\sigma)=\LB(\Pi)+\frac{e_{12}(\sigma)+e_{13}(\sigma)+e_{23}(\sigma)}{2}.
\end{equation}

\begin{theorem}[small excess]\label{thm:small-excess}
Every permutation $\sigma$ with $F_\Pi(\sigma)\leq\LB(\Pi)+1$ is a pivot solution. Consequently,
\[
\med(\Pi)\leq \LB(\Pi)+1 \iff \min_r P_r\leq \LB(\Pi)+1 \iff \exists r:\ M_r\geq \beta_r-1,
\]
and in that case $\med(\Pi)=\min_r P_r$.
\end{theorem}

\begin{proof}
Each defect is even: $\dd(\pi_i,\sigma)$, $\dd(\sigma,\pi_j)$ and $\dd_{ij}$ have the parities of $\pi_i^{-1}\sigma$, $\sigma^{-1}\pi_j$ and of their product $\pi_i^{-1}\pi_j$. By~\eqref{eq:defects}, $e_{12}+e_{13}+e_{23}\leq 2$, so at most one defect is non-zero. The two vanishing defects share an index $r$, hence $\sigma\in I(\pi_r,\pi_s)\cap I(\pi_r,\pi_t)$. The equivalences follow from \autoref{thm:lb-characterization}(3), since $P_r=\LB+g_r$.
\end{proof}

\autoref{thm:small-excess} explains an empirical observation (\autoref{sec:experiments}): on random instances the optimum is almost always at most $\LB+1$, and pivot solutions are then optimal. It is natural to ask whether some optimal median is always a pivot solution, which would give $\med(\Pi)=\min_r P_r$. This is false.

\begin{proposition}\label{prop:pivot-not-optimal}
There are instances $\Pi$ with $\med(\Pi)<\min_r P_r$, including instances in which the group generated by $\pi_1^{-1}\pi_2$ and $\pi_1^{-1}\pi_3$ acts transitively on the ground set (connected instances).
\end{proposition}

\begin{proof}
The first phenomenon comes from disjoint unions. If $\Pi=\Pi_A\uplus\Pi_B$ (componentwise disjoint union), then distances, $D_r$, $\beta_r$ and $M_r$ are additive (by the block property of \autoref{thm:biane}). Hence $\min_r P_r(\Pi)=\min_r\big(P_r(\Pi_A)+P_r(\Pi_B)\big)$, whereas $\med(\Pi)\leq\med(\Pi_A)+\med(\Pi_B)$. Take an instance $\Pi_A$ on $6$ elements whose pivot values are $P_1(\Pi_A)=6<P_2(\Pi_A)=P_3(\Pi_A)=7$, with a median of value $6$. Let $\Pi_B$ be the same instance with the roles of $\pi_1$ and $\pi_2$ exchanged. Then $\med(\Pi)\leq 12<13=\min_r P_r(\Pi)$. The explicit instance, its pivot values and a median of value $12$ are given in \autoref{app:examples}. The same appendix gives a connected instance obtained by one extra swap, with $\min_r P_r=14$ and a permutation of value $13$.
\end{proof}

In both examples the permutation of \autoref{app:examples} lies on a geodesic between $\pi_1$ and $\pi_2$, but not on two geodesics leaving a common input. Whether every instance has an optimal median on \emph{some} geodesic $I(\pi_i,\pi_j)$, remains open.

\subsection{The crossing graph}

Common lower bounds of small rank are easy to describe through pairs of elements. Fix a pivot, say $r=1$, and write $R=\tau^1_2$ (\emph{red} cycles) and $B=\tau^1_3$ (\emph{blue} cycles).

\begin{definition}\label{def:crossing-graph}
A pair $\{a,b\}$ of distinct elements is \emph{admissible} if $a$ and $b$ belong to a common red cycle and to a common blue cycle. The \emph{crossing graph} $X(R,B)$ has the admissible pairs as vertices; two admissible pairs $\{a,b\}$ and $\{c,d\}$ are adjacent if they share an element, or if $a,b,c,d$ are distinct and the two pairs cross in a common red cycle or in a common blue cycle. The pair $(R,B)$ is \emph{$2$-sparse} if every red cycle meets every blue cycle in at most two elements.
\end{definition}

For an independent set $I$ of $X(R,B)$, let $\lambda_I$ be the product of the (pairwise disjoint, hence commuting) transpositions $(a\ b)$, $\{a,b\}\in I$.

\begin{lemma}\label{lem:is-to-clb}
For every independent set $I$ of $X(R,B)$, $\lambda_I$ is a common lower bound of $R$ and $B$ with $\ellp(\lambda_I)=|I|$. Hence $M_1\geq \alpha(X(R,B))$. If $(R,B)$ is $2$-sparse, then every common lower bound is of the form $\lambda_I$ and $M_1=\alpha(X(R,B))$.
\end{lemma}

\begin{proof}
The cycles of $\lambda_I$ are the pairs of $I$ (and fixed points). Each pair lies in a red cycle and in a blue cycle; inside each cycle the pairs are pairwise disjoint and non-crossing; and a $2$-element cycle has only one cyclic order. By \autoref{thm:biane}, $\lambda_I\lea R$ and $\lambda_I\lea B$, and $\ellp(\lambda_I)=|I|$. Conversely, let $\lambda$ be a common lower bound. By \autoref{thm:biane}(i), each cycle of $\lambda$ is contained in a red cycle and in a blue cycle. Under $2$-sparseness, it therefore has at most two elements, so $\lambda=\lambda_I$ for the set $I$ of its $2$-cycles. These pairs are admissible and pairwise disjoint, and they are non-crossing in every red and blue cycle by \autoref{thm:biane}(ii). Hence $I$ is independent.
\end{proof}

\begin{example}\label{ex:shared-element}
Pairs sharing an element must be declared adjacent. Let $\pi_1=\iota$, $\pi_2=[3\ 1\ 6\ 5\ 4\ 2]$ and $\pi_3=[5\ 6\ 2\ 3\ 1\ 4]$, so that $R=(1\ 3\ 6\ 2)(4\ 5)$ and $B=(1\ 5)(2\ 6\ 4\ 3)$. The admissible pairs are $\{2,3\},\{2,6\},\{3,6\}$. They pairwise share an element, so $X(R,B)$ is a triangle. Since $\dd_{12}=4$, $\dd_{13}=4$ and $\dd_{23}=4$, we have $\beta_1=2$ and $\LB=6$. If pairs sharing an element were not declared adjacent, the three pairs would form an independent set of size $3\geq\beta_1$. Yet $M_1=1$: $R$ induces the cyclic order $(2\ 3\ 6)$ on $\{2,3,6\}$ and $B$ induces the opposite order $(2\ 6\ 3)$, so by \autoref{thm:biane}(iii) no cycle of a common lower bound contains all three elements. By \autoref{thm:lb-characterization} and \autoref{thm:small-excess}, $\med=P_1=\LB+1=7$.
\end{example}

\autoref{fig:RBcycles} shows a $2$-sparse example. Let $\pi_1=\iota$, $\pi_2=[4\ 5\ 1\ 6\ 2\ 3\ 8\ 7]$ and $\pi_3=[6\ 5\ 7\ 8\ 1\ 2\ 4\ 3]$, that is, $R=(1\ 4\ 6\ 3)(2\ 5)(7\ 8)$ and $B=(1\ 6\ 2\ 5)(3\ 7\ 4\ 8)$. Every red cycle meets every blue cycle in at most two elements. The admissible pairs are $\{1,6\}$, $\{3,4\}$, $\{2,5\}$ and $\{7,8\}$, and the only crossings are $\{1,6\}$ with $\{3,4\}$ in the red cycle $(1\ 4\ 6\ 3)$, and $\{3,4\}$ with $\{7,8\}$ in the blue cycle $(3\ 7\ 4\ 8)$. Hence $X(R,B)$ is a path on three vertices plus an isolated vertex, and $M_1=\alpha(X(R,B))=3$. Since $\dd_{12}=5$, $\dd_{13}=6$ and $\dd_{23}=5$, we have $\beta_1=3$ and $\LB=8$. By \autoref{thm:lb-characterization}, $\sigma=\pi_1(1\ 6)(2\ 5)(7\ 8)=[6\ 5\ 3\ 4\ 2\ 1\ 8\ 7]$ attains the lower bound, with $\dd(\sigma,\pi_1)=3$, $\dd(\sigma,\pi_2)=2$ and $\dd(\sigma,\pi_3)=3$.

\begin{figure}[!h]
    \centering
    \input{figures/fig_crossing}
    \caption{Left: the red cycles of $R=\pi_1^{-1}\pi_2$ and the blue cycles of $B=\pi_1^{-1}\pi_3$ for the instance in the text, drawn clockwise. Each admissible pair is drawn as a chord of the red cycle and of the blue cycle that contain it. Right: the crossing graph $X(R,B)$. The maximum independent set $\{\{1,6\},\{2,5\},\{7,8\}\}$ (green) gives the median $\sigma=(1\ 6)(2\ 5)(7\ 8)$, which attains $\LB$; the dashed pair $\{3,4\}$ is not used.}
    \label{fig:RBcycles}
\end{figure}

%% file: figures/fig_crossing.tex
\begin{tikzpicture}[scale=1.0]
  \node[font=\small, red!75!black] at (-1.0,1.25) {$R$};
  \drawcycle{Ra}{0}{0.6}{0.6}{1,4,6,3}{redc}
  \drawcycle{Rb}{1.6}{0.6}{0.4}{2,5}{redc}
  \drawcycle{Rc}{2.8}{0.6}{0.4}{7,8}{redc}
  \draw[chordsel] (Ra-1)--(Ra-6); \draw[chordoff] (Ra-3)--(Ra-4);
  \draw[chordsel] (Rb-2)--(Rb-5); \draw[chordsel] (Rc-7)--(Rc-8);
  \node[font=\small, blue!70!black] at (-1.0,-0.75) {$B$};
  \drawcycle{Ba}{0.3}{-1.3}{0.6}{1,6,2,5}{bluec}
  \drawcycle{Bb}{2.3}{-1.3}{0.6}{3,7,4,8}{bluec}
  \draw[chordsel] (Ba-1)--(Ba-6); \draw[chordsel] (Ba-2)--(Ba-5);
  \draw[chordoff] (Bb-3)--(Bb-4); \draw[chordsel] (Bb-7)--(Bb-8);
  \begin{scope}[xshift=4.6cm, yshift=-0.35cm]
    \node[gsel] (x16) at (0,0.9) {$16$};
    \node[gvert] (x34) at (1.2,0.9) {$34$};
    \node[gsel] (x78) at (1.2,-0.3) {$78$};
    \node[gsel] (x25) at (0,-0.3) {$25$};
    \draw[red!75!black, line width=1pt] (x16)--(x34) node[midway, above, font=\scriptsize]{red};
    \draw[blue!70!black, line width=1pt] (x34)--(x78) node[midway, right, font=\scriptsize]{blue};
    \node[font=\small] at (0.6,1.65) {$X(R,B)$};
  \end{scope}
\end{tikzpicture}

%% file: sections/pivot_hardness.tex
\section{Hardness of the pivot problem}
\label{sec:pivot}

We now show that computing $M_r$, the length of the longest walk from an input towards both other inputs, is hard. We state the problem with pivot $\pi_1$.

\defproblema
{Common Geodesic Prefix}
{Three permutations $\pi_1,\pi_2,\pi_3$ of the same size and a non-negative integer $k$.}
{Is there a permutation $\sigma\in I(\pi_1,\pi_2)\cap I(\pi_1,\pi_3)$ with $\dd(\pi_1,\sigma)\geq k$? Equivalently, is $M_1\geq k$?}

By \autoref{lem:pivot-value}, this is the same as asking for a pivot solution at $\pi_1$ of median value at most $D_1-k$. The problem belongs to \NP: membership of $\sigma$ in an interval is checked by computing three distances. In the maximization version, {\sc Max Common Geodesic Prefix}, the goal is to find such a $\sigma$ maximizing $\dd(\pi_1,\sigma)$.

\paragraph*{$2$-subdivisions.}
For a graph $G$, the $2$-subdivision $\Subd(G)$ is obtained by replacing every edge $uv$ by a path $u,x_{uv},x_{vu},v$ of length three. We use Poljak's identity~\cite{zbMATH03445275} in the following constructive form.

\begin{lemma}\label{lem:poljak}
For every graph $G$ with $m$ edges, $\alpha(\Subd(G))=\alpha(G)+m$. Moreover, from an independent set $S$ of $G$ one obtains in polynomial time an independent set of $\Subd(G)$ of size $|S|+m$, and from an independent set $I$ of $\Subd(G)$ one obtains in polynomial time an independent set of $G$ of size at least $|I|-m$.
\end{lemma}

\begin{proof}
Given $S$, add for every edge $uv$ one subdivision vertex: $x_{vu}$ if $u\in S$, $x_{uv}$ if $v\in S$, and either one otherwise (at most one endpoint is in $S$). Conversely, given $I$, let $S=I\cap V(G)$ and let $b$ be the number of edges with both endpoints in $S$. For each such edge, neither $x_{uv}$ nor $x_{vu}$ is in $I$; every other edge contributes at most one subdivision vertex. Hence $|I|\leq |S|+m-b$. Removing one endpoint of each edge inside $S$ yields an independent set of $G$ of size at least $|S|-b\geq |I|-m$.
\end{proof}

\paragraph*{The construction.}
The following construction represents $\Subd(G)$ as a crossing graph (see \autoref{fig:subdiv2}).

\begin{construction}\label{constr:subdivision}
Let $G$ be a simple graph. For every vertex $z$ of $\Subd(G)$ introduce two elements $z^-,z^+$, and let $s_z=\{z^-,z^+\}$. For every vertex $v$ of $G$ with neighbors $u_1,\ldots,u_d$, create the red cycle
\[
\bigl(v^-,\ x_{vu_1}^-,\ x_{vu_2}^-,\ \ldots,\ x_{vu_d}^-,\ v^+,\ x_{vu_d}^+,\ \ldots,\ x_{vu_2}^+,\ x_{vu_1}^+\bigr),
\]
and the blue cycle $(v^-\ v^+)$. For every edge $uv$ of $G$ create the blue cycle $(x_{uv}^-\ x_{vu}^-\ x_{uv}^+\ x_{vu}^+)$. Let $R$ and $B$ be the permutations with these red and blue cycles, and let $\Pi_G=(\iota,R,B)$.
\end{construction}

\begin{example}\label{ex:subdivision}
Let $G$ be the path $u\,v\,w$. Then $\Subd(G)$ is the path $u,x_{uv},x_{vu},v,x_{vw},x_{wv},w$, with $\alpha(\Subd(G))=\alpha(G)+m=2+2=4$. Number the $14$ elements as $u^\mp=1,2$, $x_{uv}^\mp=3,4$, $v^\mp=5,6$, $x_{vu}^\mp=7,8$, $x_{vw}^\mp=9,10$, $w^\mp=11,12$ and $x_{wv}^\mp=13,14$. \autoref{constr:subdivision} gives the red cycles $(1\ 3\ 2\ 4)$, $(5\ 7\ 9\ 6\ 10\ 8)$ and $(11\ 13\ 12\ 14)$, and the blue cycles $(1\ 2)$, $(5\ 6)$, $(11\ 12)$, $(3\ 7\ 4\ 8)$ and $(9\ 13\ 10\ 14)$. The three input permutations are therefore
\[
\pi_1=\iota_{14},\quad
\pi_2=[3\ 4\ 2\ 1\ 7\ 10\ 9\ 5\ 6\ 8\ 13\ 14\ 12\ 11],\quad
\pi_3=[2\ 1\ 7\ 8\ 6\ 5\ 4\ 3\ 13\ 14\ 12\ 11\ 10\ 9].
\]
The independent set $\{u,x_{vu},x_{vw},w\}$ of $\Subd(G)$ gives the common lower bound $\lambda=(1\ 2)(7\ 8)(9\ 10)(11\ 12)$ of rank $4=M_1$ (\autoref{fig:subdiv2}). Here $\dd_{12}=11$, $\dd_{13}=9$ and $\dd_{23}=12$, so $\beta_1=4=M_1$: since $G$ is a tree, the lower bound $\LB=16$ is attained by $\sigma=\lambda$.
\end{example}

\begin{figure}[!h]
    \centering
    \input{figures/fig_subdiv}
    \caption{\autoref{constr:subdivision} for the path $G=u\,v\,w$ (\autoref{ex:subdivision}). (a)~$G$ and $\Subd(G)$. (b)~The red cycles, one per vertex of $G$: in the red cycle of $v$, the chord $s_v=\{5,6\}$ crosses the chords $s_{x_{vu}}=\{7,8\}$ and $s_{x_{vw}}=\{9,10\}$, which are nested. (c)~The blue cycles: one $2$-cycle per vertex of $G$ and one $4$-cycle per edge, in which the two subdivision chords cross. Every chord lies in exactly one red and one blue cycle, and two chords cross somewhere if and only if the corresponding vertices are adjacent in $\Subd(G)$. The independent set $\{u,x_{vu},x_{vw},w\}$ and its chords are shown in green.}
    \label{fig:subdiv2}
\end{figure}

\begin{lemma}\label{lem:subdivision-crossing}
For every simple graph $G$, the pair $(R,B)$ of \autoref{constr:subdivision} is $2$-sparse, its admissible pairs are exactly the pairs $s_z$, and $X(R,B)\cong\Subd(G)$ via $s_z\mapsto z$. Consequently, $M_1(\Pi_G)=\alpha(G)+|E(G)|$.
\end{lemma}

\begin{proof}
The red cycle of $v$ meets the blue cycle $(v^-\ v^+)$ in $\{v^-,v^+\}$, meets the blue cycle of an edge $vu$ in $\{x_{vu}^-,x_{vu}^+\}$, and is disjoint from all other blue cycles. Hence $(R,B)$ is $2$-sparse and the admissible pairs are the pairs $s_z$; they are pairwise disjoint, so adjacency in $X(R,B)$ only comes from crossings. In the red cycle of $v$, the chord $s_v$ separates the elements $x^-_{vu_i}$ from the elements $x^+_{vu_i}$ and thus crosses every chord $s_{x_{vu_i}}$, while for $i<j$ the elements appear in the order $x^-_{vu_i},x^-_{vu_j},x^+_{vu_j},x^+_{vu_i}$, so $s_{x_{vu_i}}$ and $s_{x_{vu_j}}$ do not cross. In the blue cycle of an edge $uv$, the chords $s_{x_{uv}}$ and $s_{x_{vu}}$ alternate, hence cross; the blue $2$-cycles contain a single chord. Therefore $s_z$ and $s_{z'}$ are adjacent exactly when $zz'$ is an edge of $\Subd(G)$. The last statement follows from \autoref{lem:is-to-clb} and \autoref{lem:poljak}.
\end{proof}

All blue cycles have length at most $4$, and the red cycle of $v$ has length $2(\deg(v)+1)$.

\begin{theorem}\label{thm:cgp-hard}
{\sc Common Geodesic Prefix} is \NP-complete, and {\sc Max Common Geodesic Prefix} is \APX-hard. Both statements hold even when every cycle of $\pi_1^{-1}\pi_2$ has length at most $8$ and every cycle of $\pi_1^{-1}\pi_3$ has length at most $4$.
\end{theorem}

\begin{proof}
\NP-hardness: by \autoref{lem:subdivision-crossing}, $G$ has an independent set of size $k'$ if and only if $M_1(\Pi_G)\geq k'+|E(G)|$. {\sc Maximum Independent Set} is \NP-hard on graphs of maximum degree three~\cite{garey1976simplified}, which gives red cycles of length at most $8$.

\APX-hardness: we give an L-reduction~\cite{papadimitriou1991optimization} from {\sc Maximum Independent Set} on cubic graphs, which is \APX-complete~\cite{alimonti2000some}. Let $G$ be cubic with $n$ vertices and $m=3n/2$ edges. Since every vertex has degree $3$, a greedy algorithm finds an independent set of size at least $n/4$, so $\alpha(G)\geq n/4$ and $m\leq 6\alpha(G)$. Hence $M_1(\Pi_G)=\alpha(G)+m\leq 7\,\alpha(G)$. Conversely, let $\sigma$ be a feasible solution with $\ellp(\lambda)=h$, where $\lambda=\pi_1^{-1}\sigma$. By $2$-sparseness, $\lambda=\lambda_I$ for an independent set $I$ of $X(R,B)\cong\Subd(G)$ with $|I|=h$, and \autoref{lem:poljak} yields an independent set $S$ of $G$ with $|S|\geq h-m$. Therefore $\alpha(G)-|S|\leq (\alpha(G)+m)-h=M_1(\Pi_G)-h$. Both conditions of an L-reduction hold, with constants $7$ and $1$.
\end{proof}

\begin{corollary}\label{cor:pivot-value-hard}
Given $\Pi$, a pivot $r$ and an integer $f$, it is \NP-complete to decide whether $P_r\leq f$, that is, whether some pivot solution at $\pi_r$ has median value at most $f$.
\end{corollary}

\begin{remark}
\autoref{thm:cgp-hard} also shows why the median lower bound cannot be used directly as the threshold of a reduction from independent sets on $\Subd(G)$. For the instances $\Pi_G$, the Gromov product $\beta_1=(3m+c(R^{-1}B))/2$ is fixed by the construction. For instance, $\beta_1=6$ for $G=K_3$ and $\beta_1=9$ for $G=C_5$, whereas $\alpha(\Subd(G))$ equals $4$ and $7$, respectively. In both cases $g_1>0$, so these instances never attain the median lower bound.
\end{remark}

%% file: figures/fig_subdiv.tex
\begin{tikzpicture}[scale=1.0]
  \begin{scope}[yshift=0.2cm]
  \node[font=\small] at (-0.4,1.55) {(a)};
  \node[gvert] (gu) at (0.2,1.5) {$u$}; \node[gvert] (gv) at (1.2,1.5) {$v$}; \node[gvert] (gw) at (2.2,1.5) {$w$};
  \draw (gu)--(gv)--(gw); \node[font=\scriptsize] at (3.0,1.5) {$G$};
  \node[gsel] (su) at (0,0.4) {$u$};
  \node[gvert] (s1) at (0.55,0.4) {}; \node[gsel] (s2) at (1.1,0.4) {};
  \node[gvert] (sv) at (1.65,0.4) {$v$};
  \node[gsel] (s3) at (2.2,0.4) {}; \node[gvert] (s4) at (2.75,0.4) {};
  \node[gsel] (sw) at (3.3,0.4) {$w$};
  \draw (su)--(s1)--(s2)--(sv)--(s3)--(s4)--(sw);
  \node[font=\tiny, above=1pt] at (s1.north) {$x_{uv}$}; \node[font=\tiny, above=1pt] at (s2.north) {$x_{vu}$};
  \node[font=\tiny, above=1pt] at (s3.north) {$x_{vw}$}; \node[font=\tiny, above=1pt] at (s4.north) {$x_{wv}$};
  \node[font=\scriptsize] at (1.65,-0.35) {$\Subd(G)$};
  \end{scope}
  \begin{scope}[xshift=5.6cm]
  \node[font=\small] at (-0.9,1.55) {(b)};
  \drawcycle{Ru}{0}{0.6}{0.5}{1,3,2,4}{redc}
  \drawcycle{Rv}{1.75}{0.6}{0.75}{5,7,9,6,10,8}{redc}
  \drawcycle{Rw}{3.5}{0.6}{0.5}{11,13,12,14}{redc}
  \draw[chordsel] (Ru-1)--(Ru-2); \draw[chord] (Ru-3)--(Ru-4);
  \draw[chord] (Rv-5)--(Rv-6); \draw[chordsel] (Rv-7)--(Rv-8); \draw[chordsel] (Rv-9)--(Rv-10);
  \draw[chordsel] (Rw-11)--(Rw-12); \draw[chord] (Rw-13)--(Rw-14);
  \end{scope}
  \begin{scope}[xshift=5.4cm, yshift=-1.6cm]
  \node[font=\small] at (-0.9,0.75) {(c)};
  \drawcycle{Bu}{-0.35}{0}{0.32}{1,2}{bluec}
  \drawcycle{Be}{0.85}{0}{0.5}{3,7,4,8}{bluec}
  \drawcycle{Bv}{2.05}{0}{0.32}{5,6}{bluec}
  \drawcycle{Bf}{3.25}{0}{0.5}{9,13,10,14}{bluec}
  \drawcycle{Bw}{4.45}{0}{0.32}{11,12}{bluec}
  \draw[chordsel] (Bu-1)--(Bu-2); \draw[chord] (Bv-5)--(Bv-6); \draw[chordsel] (Bw-11)--(Bw-12);
  \draw[chord] (Be-3)--(Be-4); \draw[chordsel] (Be-7)--(Be-8);
  \draw[chordsel] (Bf-9)--(Bf-10); \draw[chord] (Bf-13)--(Bf-14);
  \end{scope}
\end{tikzpicture}

%% file: sections/closest.tex
\section{\NP-completeness of {\sc Swap Closest} for three permutations}
\label{sec:closest}

The closest objective imposes a separate budget on each of the three distances. We exploit this by padding an instance of {\sc Common Geodesic Prefix} so that two triangle inequalities become tight at the threshold. This forces every feasible solution to be a pivot solution, while the third distance keeps some slack.

\begin{construction}\label{constr:closest}
Let $\Pi=(\pi_1,\pi_2,\pi_3)$ be permutations of size $N$ and let $k$ be an integer with $0\leq k\leq\beta_1(\Pi)$. Set
\[
x=\max\{0,\ \dd_{12}-2k,\ \dd_{13}-2k\},\qquad y=x+2k-\dd_{12},\qquad z=x+2k-\dd_{13},\qquad f=x+k,
\]
so that $x,y,z\geq 0$, and define
\[
\begin{array}{rcl}
\pi'_1 &=& \pi_1 \uplus \rho_{x+1} \uplus \iota_{y+1} \uplus \iota_{z+1},\\[1mm]
\pi'_2 &=& \pi_2 \uplus \iota_{x+1} \uplus \rho_{y+1} \uplus \iota_{z+1},\\[1mm]
\pi'_3 &=& \pi_3 \uplus \iota_{x+1} \uplus \iota_{y+1} \uplus \rho_{z+1}.
\end{array}
\]
Let $\Pi'_k=(\pi'_1,\pi'_2,\pi'_3)$. We call the four parts of the ground set the blocks $A$, $X$, $Y$, $Z$.
\end{construction}

\begin{figure}[!h]
\centering
\begin{tikzpicture}[scale=0.82, every node/.style={font=\footnotesize}]
  \def\h{0.55}
  \foreach \i/\lab in {0/$\pi'_1$,1/$\pi'_2$,2/$\pi'_3$}{
    \node[anchor=east] at (-0.1,-\i*\h-0.25) {\lab};
  }
  \foreach \i/\t in {0/$\pi_1$,1/$\pi_2$,2/$\pi_3$}{
    \draw[fill=gray!15] (0,-\i*\h) rectangle (3,-\i*\h-0.5); \node at (1.5,-\i*\h-0.25) {\t};
  }
  \foreach \i/\t/\c in {0/$\rho_{x+1}$/red!20,1/$\iota_{x+1}$/white,2/$\iota_{x+1}$/white}{
    \draw[fill=\c] (3.2,-\i*\h) rectangle (5.0,-\i*\h-0.5); \node at (4.1,-\i*\h-0.25) {\t};
  }
  \foreach \i/\t/\c in {0/$\iota_{y+1}$/white,1/$\rho_{y+1}$/red!20,2/$\iota_{y+1}$/white}{
    \draw[fill=\c] (5.2,-\i*\h) rectangle (7.0,-\i*\h-0.5); \node at (6.1,-\i*\h-0.25) {\t};
  }
  \foreach \i/\t/\c in {0/$\iota_{z+1}$/white,1/$\iota_{z+1}$/white,2/$\rho_{z+1}$/red!20}{
    \draw[fill=\c] (7.2,-\i*\h) rectangle (9.0,-\i*\h-0.5); \node at (8.1,-\i*\h-0.25) {\t};
  }
  \node at (1.5,0.3) {block $A$}; \node at (4.1,0.3) {block $X$}; \node at (6.1,0.3) {block $Y$}; \node at (8.1,0.3) {block $Z$};
\end{tikzpicture}
\caption{\autoref{constr:closest}. Block $X$ adds $x$ to the distances from $\pi'_1$ to $\pi'_2$ and to $\pi'_3$, block $Y$ adds $y$ to the distances from $\pi'_2$, and block $Z$ adds $z$ to the distances from $\pi'_3$. The values are chosen so that $\dd(\pi'_1,\pi'_2)=\dd(\pi'_1,\pi'_3)=2f$.}
\label{fig:closest-padding}
\end{figure}

\begin{lemma}\label{lem:closest-distances}
In \autoref{constr:closest}, $\dd(\pi'_1,\pi'_2)=\dd(\pi'_1,\pi'_3)=2f$ and $\dd(\pi'_2,\pi'_3)=2f-2(\beta_1-k)\leq 2f$. In particular, $\LB_{\clo}(\Pi'_k)=f$.
\end{lemma}

\begin{proof}
By additivity over disjoint unions and $\dd(\iota_t,\rho_t)=t-1$,
$\dd(\pi'_1,\pi'_2)=\dd_{12}+x+y=\dd_{12}+2x+2k-\dd_{12}=2f$ and, symmetrically, $\dd(\pi'_1,\pi'_3)=\dd_{13}+x+z=2f$. Finally,
$\dd(\pi'_2,\pi'_3)=\dd_{23}+y+z=\dd_{23}+2x+4k-\dd_{12}-\dd_{13}=2x+4k-2\beta_1=2f-2(\beta_1-k)$.
\end{proof}

\begin{theorem}\label{thm:closest-equivalence}
Let $\Pi$ and $0\leq k\leq\beta_1(\Pi)$ be given, and let $\Pi'_k$ and $f$ be as in \autoref{constr:closest}. Then $\clo(\Pi'_k)\leq f$ if and only if $M_1(\Pi)\geq k$.
\end{theorem}

\begin{proof}
($\Leftarrow$) Since intervals of the absolute order are graded, there is a common lower bound $\lambda$ of $\tau_2=\pi_1^{-1}\pi_2$ and $\tau_3=\pi_1^{-1}\pi_3$ with $\ellp(\lambda)=k$. Let $\sigma=\pi_1\lambda$ and $\sigma'=\sigma\uplus\iota_{x+1}\uplus\iota_{y+1}\uplus\iota_{z+1}$. Then $\dd(\sigma',\pi'_1)=k+x=f$, $\dd(\sigma',\pi'_2)=(\dd_{12}-k)+y=f$ and $\dd(\sigma',\pi'_3)=(\dd_{13}-k)+z=f$.

($\Rightarrow$) Let $\sigma'$ satisfy $\dd(\sigma',\pi'_i)\leq f$ for $i=1,2,3$. By \autoref{lem:closest-distances}, $2f=\dd(\pi'_1,\pi'_2)\leq\dd(\pi'_1,\sigma')+\dd(\sigma',\pi'_2)\leq 2f$, so both inequalities are tight and $\sigma'\in I(\pi'_1,\pi'_2)$; similarly $\sigma'\in I(\pi'_1,\pi'_3)$, and $\dd(\pi'_1,\sigma')=f$. Thus $\lambda'=\pi_1'^{-1}\sigma'$ is a common lower bound of $\pi_1'^{-1}\pi'_2$ and $\pi_1'^{-1}\pi'_3$ of rank $f$. These two permutations map each block onto itself; on the blocks $A,X,Y,Z$ they are respectively $(\tau_2,\tau_3)$, $(\rho_{x+1}^{-1},\rho_{x+1}^{-1})$, $(\rho_{y+1},\iota)$ and $(\iota,\rho_{z+1})$. By the block property of \autoref{thm:biane}, $\lambda'=\lambda_A\uplus\lambda_X\uplus\lambda_Y\uplus\lambda_Z$ where each part is a common lower bound inside its block. Since $\lambda_Y\lea\iota$ and $\lambda_Z\lea\iota$, both are identities, and $\ellp(\lambda_X)\leq\ellp(\rho_{x+1}^{-1})=x$. Hence $\ellp(\lambda_A)=f-\ellp(\lambda_X)\geq f-x=k$, and $\lambda_A$ is a common lower bound of $\tau_2$ and $\tau_3$, so $M_1(\Pi)\geq k$.
\end{proof}

The size of $\Pi'_k$ is $N+x+y+z+3=O(N)$, since all distances are at most $N$. Since $M_1(\Pi)\leq\beta_1(\Pi)$ always holds, a value $k>\beta_1$ is a trivial no-instance of {\sc Common Geodesic Prefix}. Combined with \autoref{thm:cgp-hard}, we obtain the following.

\begin{theorem}\label{thm:swap-closest-npc}
{\sc Swap Closest} is \NP-complete for three input permutations. Moreover, it is \NP-complete to decide whether $\clo(\Pi)=\LB_{\clo}(\Pi)$, even when restricted to instances in which the largest pairwise distance is even and attained by two of the three pairs.
\end{theorem}

\begin{proof}
Membership in \NP is clear. Given an instance $(\Pi,k)$ of {\sc Common Geodesic Prefix}, output a fixed no-instance if $k>\beta_1(\Pi)$ and $(\Pi'_k,f)$ otherwise. By \autoref{thm:closest-equivalence} this is a polynomial reduction, and by \autoref{lem:closest-distances}, $f=\LB_{\clo}(\Pi'_k)$, with the largest pairwise distance $2f$ attained by the pairs $\{1,2\}$ and $\{1,3\}$.
\end{proof}

Without padding, the same argument proves hardness of a natural variant in which each input has its own radius. Indeed, $\dd(\sigma,\pi_1)\leq k$, $\dd(\sigma,\pi_2)\leq\dd_{12}-k$ and $\dd(\sigma,\pi_3)\leq \dd_{13}-k$ hold if and only if $\sigma$ is a pivot solution at $\pi_1$ with $\dd(\pi_1,\sigma)=k$.

\begin{corollary}\label{cor:radii}
Given three permutations and radii $r_1,r_2,r_3$, deciding whether some permutation $\sigma$ satisfies $\dd(\sigma,\pi_i)\leq r_i$ for $i=1,2,3$ is \NP-complete.
\end{corollary}

\begin{remark}[Why this does not extend to {\sc Swap Median}]
A median attaining $\LB$ must make \emph{all three} triangle inequalities tight (\autoref{obs:lower-bound-paths}), that is, it must be a pivot solution of rank exactly $\beta_r$. For a disjoint union, pivot gaps add up, $g_r(\Pi\uplus\Pi'')=g_r(\Pi)+g_r(\Pi'')$, and they are non-negative. Hence padding blocks cannot compensate for the slack $\beta_1-k$ of the third pair, which is precisely what \autoref{constr:closest} leaves free. For this reason, the next section proves the hardness of {\sc Swap Median} by a different route, which does not use the lower bound.
\end{remark}

%% file: sections/median.tex
\section{\NP-completeness of {\sc Swap Median} for three permutations}
\label{sec:median}

The lower-bound approach of \autoref{sec:closest} cannot work for the median, because a median attaining $\LB$ must make all three triangle inequalities tight. We therefore use a different idea. We take two inputs that differ by disjoint swaps. Then the interval between them is a Boolean cube, and we show that some optimal median always lies in this cube. Optimizing over the cube turns out to be equivalent to finding a maximum cycle decomposition of an Eulerian digraph, which is \NP-hard by a classical result of Colbourn on partial Latin squares~\cite{colbourn1984complexity}.

\subsection{Two inputs differing by disjoint swaps}

Let $\tau=t_1t_2\cdots t_k$ be a product of $k$ pairwise disjoint transpositions, so $\ellp(\tau)=k$. For $S\subseteq[k]$ write $\tau_S=\prod_{i\in S}t_i$. Since $\tau$ has cycles of length at most two, \autoref{thm:biane} gives $[\iota,\tau]=\{\tau_S : S\subseteq[k]\}$. In other words, the geodesics between $\iota$ and $\tau$ form a Boolean cube.

We use a classical formula of Cohn and Lempel~\cite{cohn1972cycle}. Let $x$ be a cyclic permutation of a set $U$, and let $t_1,\ldots,t_m$ be disjoint transpositions with all their points in $U$. The \emph{interlace matrix} $A$ is the $m\times m$ matrix over $\mathrm{GF}(2)$ with $A_{ij}=1$ if and only if $i\neq j$ and the two pairs of $t_i$ and $t_j$ cross on the cycle $x$.

\begin{theorem}[Cohn and Lempel~\cite{cohn1972cycle}]\label{thm:cohn-lempel}
With the notation above, the restriction of $x\,t_1\cdots t_m$ to $U$ has exactly $1+m-\mathrm{rank}_{\mathrm{GF}(2)}(A)$ cycles.
\end{theorem}

The next lemma is the key structural fact. It says that, from any permutation, one can reach the cube $[\iota,\tau]$ by paying at most the defect of the permutation with respect to the pair $(\iota,\tau)$.

For example, let $x=(1\ 2\ 3\ 4)$. With the single chord $t_1=(1\ 3)$ we get $x\,t_1=(1\ 4)(2\ 3)$, that is, $2=1+1-0$ cycles. With the two crossing chords $t_1=(1\ 3)$ and $t_2=(2\ 4)$ the interlace matrix has rank $2$, and indeed $x\,t_1t_2=(1\ 4\ 3\ 2)$ has $1=1+2-2$ cycle. Geometrically, a chord splits the cycle into two arcs, and a second chord crossing it merges them again.

\begin{lemma}[projection onto the cube]\label{lem:cube-projection}
Let $\tau$ be a product of $k$ disjoint transpositions. For every $\sigma\in S_n$ there is $S\subseteq[k]$ with
\[
\dd(\sigma,\tau_S)\ \leq\ \ellp(\sigma)+\ellp(\sigma^{-1}\tau)-k.
\]
\end{lemma}

\begin{proof}
Let $x=\sigma^{-1}$. Since $\dd(\sigma,\tau_S)=n-c(x\tau_S)$ and $c(\sigma)=c(\sigma^{-1})$, the claim is equivalent to
\begin{equation}\label{eq:cube}
\max_{S\subseteq[k]} c(x\tau_S)\ \geq\ c(x)+c(x\tau)+k-n .
\end{equation}
Let $G$ be the group generated by $x$ and $t_1,\ldots,t_k$. Every transposition has both of its points in a single orbit of $G$, and every $x\tau_S$ maps each orbit onto itself. Hence both sides of~\eqref{eq:cube} are sums over the orbits. It suffices to prove, for each orbit $O$ with $n_O$ points and $k_O$ transpositions,
\begin{equation}\label{eq:cube-orbit}
\max_{S} c_O(x\tau_S)\ \geq\ c_O(x)+c_O(x\tau)+k_O-n_O ,
\end{equation}
where $c_O$ counts cycles contained in $O$ and $S$ ranges over subsets of the $k_O$ transpositions of $O$. If $k_O=0$, then $O$ is a single cycle of $x$, and~\eqref{eq:cube-orbit} reads $1\geq 2-n_O$, which holds.

Assume $k_O\geq 1$. We first find $W$ such that $x\tau_W$ is a single cycle on $O$. Start with $W=\emptyset$. If $y=x\tau_W$ has at least two cycles on $O$, some transposition $t$ of $O$ has its points in two different cycles of $y$. Otherwise every cycle of $y$ would be invariant under all $t_i$, hence under $x=y\tau_W$, hence under $G$, contradicting the fact that $O$ is a single orbit. Replacing $W$ by $W\triangle\{t\}$ multiplies $y$ by $t$ on the right, which merges these two cycles (\autoref{lem:swap-effect}). Repeating this, we reach a $W$ such that $x'=x\tau_W$ is a single cycle on $O$.

Now $x\tau_S=x'\tau_{S\triangle W}$ for every $S$. Let $A$ be the interlace matrix of the transpositions of $O$ with respect to $x'$, and for a set $Z$ of transpositions let $r(Z)=\mathrm{rank}(A[Z])$. By \autoref{thm:cohn-lempel}, $c_O(x\tau_S)=1+|S\triangle W|-r(S\triangle W)$. With $\overline{W}$ the complement of $W$ among the transpositions of $O$, this gives $c_O(x)=1+|W|-r(W)$ and $c_O(x\tau)=1+|\overline W|-r(\overline W)$. Since $n_O\geq 2k_O$, the right-hand side of~\eqref{eq:cube-orbit} is at most $2-r(W)-r(\overline W)$. If $r(W)+r(\overline W)\geq1$, the bound is at most $1$, and~\eqref{eq:cube-orbit} holds because every permutation has at least one cycle on $O$. Otherwise $A[W]=0$ and $A[\overline W]=0$. If $W\neq\emptyset$, choosing $S=\emptyset$ gives $c_O(x)=1+|W|\geq 2$. If $W=\emptyset$, choosing $S$ as the set of all transpositions of $O$ gives $c_O(x\tau)=1+k_O\geq 2$. In both cases~\eqref{eq:cube-orbit} holds.
\end{proof}

\begin{theorem}\label{thm:median-cube}
Let $\Pi=(\pi_1,\pi_2,\pi_3)$ be an instance such that $\tau=\pi_1^{-1}\pi_2$ is a product of $k$ disjoint transpositions, and let $\beta=\pi_1^{-1}\pi_3$. Then some optimal median lies in $I(\pi_1,\pi_2)=\{\pi_1\tau_S\}$, and
\[
\med(\Pi)\ =\ k+n-\max_{S\subseteq[k]} c(\tau_S\beta).
\]
\end{theorem}

\begin{proof}
By left-invariance we may assume $\pi_1=\iota$. For $\sigma=\tau_S$ we get $F_\Pi(\tau_S)=|S|+(k-|S|)+\dd(\tau_S,\beta)=k+n-c(\tau_S\beta)$, because $\tau_S^{-1}=\tau_S$. Conversely, let $\sigma$ be arbitrary and let $e=\ellp(\sigma)+\ellp(\sigma^{-1}\tau)-k$. Then $F_\Pi(\sigma)=k+e+\dd(\sigma,\beta)$. Take $S$ as in \autoref{lem:cube-projection}. By the triangle inequality, $F_\Pi(\tau_S)=k+\dd(\tau_S,\beta)\leq k+\dd(\tau_S,\sigma)+\dd(\sigma,\beta)\leq k+e+\dd(\sigma,\beta)=F_\Pi(\sigma)$.
\end{proof}

\subsection{Transition systems of Eulerian digraphs}

The quantity $\max_S c(\tau_S\beta)$ has a classical graph-theoretic meaning. Let $D$ be an Eulerian digraph (loops and parallel arcs allowed) in which every vertex has in-degree and out-degree $1$ or $2$. A \emph{transition system} chooses, at each vertex, a bijection from its in-arcs to its out-arcs; following these bijections decomposes the arc set into closed trails. Splitting a closed trail at a repeated vertex increases the number of trails. Hence the maximum number of closed trails over all transition systems equals the maximum number of cycles in a decomposition of the arc set of $D$ into arc-disjoint directed cycles.

\begin{lemma}\label{lem:transition}
Let $D$ be such a digraph with arc set $[n]$. Fix a default transition system and let $\beta(e)$ be the arc that follows $e$ in it. For every vertex $v$ with out-arcs $f,f'$, let $t_v=(f\ f')$. The transpositions $t_v$ are pairwise disjoint. Moreover, for every set $S$ of vertices, the cycles of $\tau_S\beta$ are exactly the closed trails of the transition system obtained from the default one by switching at the vertices of $S$.
\end{lemma}

\begin{proof}
Every arc is an out-arc of exactly one vertex, so the transpositions are disjoint and $\beta$ is a permutation. If $e$ enters $v$, then $\tau_S\beta(e)$ is the default successor of $e$ when $v\notin S$, and the other out-arc of $v$ when $v\in S$.
\end{proof}

\subsection{The reduction}

{\sc Partial Latin Square Completion} asks whether an $m\times m$ array, with some cells filled with symbols from $[m]$ and no symbol repeated in a row or a column, can be completed to a Latin square. Colbourn~\cite{colbourn1984complexity} proved that this problem is \NP-complete. We use it instead of Caprara's theorem that the maximum cycle decomposition of undirected Eulerian graphs is \NP-hard~\cite{caprara1999sorting}, because \autoref{lem:transition} requires \emph{directed} graphs: at a vertex with two in-arcs and two out-arcs there are exactly two transitions, matching the two choices of a transposition, whereas an undirected vertex of degree four has three. The reduction has four steps:
\[
L\ \longmapsto\ D_L\ \longmapsto\ D'_L\ \longmapsto\ (\beta,\tau)\ \longmapsto\ \Pi=(\iota,\tau,\beta).
\]

\paragraph*{Step 1: the digraph of a partial Latin square.}
Following Colbourn, we associate with $L$ a tripartite digraph $D_L$. Its vertices are the rows $r_i$, the columns $c_j$ and the symbols $s_\ell$. It has an arc $r_i\to c_j$ for every empty cell $(i,j)$, an arc $c_j\to s_\ell$ whenever $\ell$ does not occur in column $j$, and an arc $s_\ell\to r_i$ whenever $\ell$ does not occur in row $i$ (\autoref{fig:latin}(b)).

\begin{lemma}\label{lem:latin}
$D_L$ is Eulerian, every directed cycle of $D_L$ has length at least $3$, and $L$ is completable if and only if the arcs of $D_L$ can be decomposed into $|A(D_L)|/3$ directed cycles.
\end{lemma}

\begin{proof}
At a row vertex, the out-degree is the number of empty cells of the row and the in-degree is the number of symbols missing from the row; these numbers are equal. The same argument applies to columns. At a symbol vertex, both degrees are equal to $m$ minus the number of occurrences of the symbol. Arcs always go from rows to columns, from columns to symbols, and from symbols to rows, so every directed cycle has length divisible by~$3$. Hence a decomposition into $|A(D_L)|/3$ cycles consists of directed triangles $r_i\to c_j\to s_\ell\to r_i$. Such a triangle means that symbol $\ell$ is written in cell $(i,j)$: the arc $r_i\to c_j$ is used once because the cell is filled once, and the arcs $c_j\to s_\ell$ and $s_\ell\to r_i$ are used once because $\ell$ enters column $j$ and row $i$ exactly once. Thus a decomposition into triangles is exactly a completion of $L$~\cite{colbourn1984complexity}.
\end{proof}

\paragraph*{Step 2: reducing the degrees to two.}
A vertex of $D_L$ may have in- and out-degree $d>2$, and then it has $d!$ transitions, which cannot be encoded by a single transposition. Replace every vertex $v$ with $d\geq 3$ by a \emph{bubble-sort network} (\autoref{fig:latin}(c)). This network has $d$ input wires, which receive the in-arcs of $v$, and $d$ output wires, which become its out-arcs. It consists of $d(d-1)/2$ switches placed at the positions of the comparisons of bubble sort. Each switch is a vertex with two incoming and two outgoing arcs, and its two transitions either let the two wires go straight through or cross them. Every setting of the switches realizes a bijection from inputs to outputs. Conversely, every bijection is realized: running bubble sort on the target positions performs a sequence of adjacent exchanges at exactly these positions. The network is acyclic, so it creates no closed trail by itself, and the transition systems of $D_L$ correspond to the switch settings of the resulting digraph $D'_L$, with the same closed trails. Therefore the maximum number of closed trails of $D'_L$ equals that of $D_L$. All vertices of $D'_L$ have in- and out-degree $1$ or $2$, and $D'_L$ has $O(m^3)$ arcs, since each of the $3m$ vertices of $D_L$ has degree at most $m$.

\paragraph*{Steps 3 and 4: from the digraph to three permutations.}
Number the arcs of $D'_L$ by $1,\ldots,n$, fix a default transition at each vertex, and apply \autoref{lem:transition}. This gives a permutation $\beta$ of $[n]$ (the default successor of each arc) and $k$ disjoint transpositions, one for each vertex with two out-arcs. The instance is $\Pi=(\iota,\tau,\beta)$ with $\tau=t_1\cdots t_k$.

\begin{figure}[!h]
    \centering
    \resizebox{\textwidth}{!}{\input{figures/fig_latin}}
    \caption{(a)~The empty $2\times2$ square $L$ and one of its two completions. (b)~The digraph $D_L$ with its twelve arcs numbered; the completion in~(a) corresponds to the four directed triangles $\{1,5,9\}$, $\{2,8,10\}$, $\{3,6,12\}$ and $\{4,7,11\}$. (c)~The bubble-sort network used to replace a vertex of in- and out-degree $3$; each vertical segment is a switch, that is, a vertex with two in-arcs and two out-arcs.}
    \label{fig:latin}
\end{figure}

\begin{example}\label{ex:latin}
Let $L$ be the empty $2\times 2$ square. All vertices of $D_L$ have in- and out-degree $2$, so no network is needed and $D'_L=D_L$ (\autoref{fig:latin}(b)). At each vertex, let the default transition send the smaller in-arc to the smaller out-arc. This gives
\[
\beta=[5\ 7\ 6\ 8\ 9\ 10\ 11\ 12\ 1\ 2\ 3\ 4]=(1\ 5\ 9)(2\ 7\ 11\ 3\ 6\ 10)(4\ 8\ 12),
\]
and the six vertices give $\tau=(1\ 2)(3\ 4)(5\ 6)(7\ 8)(9\ 11)(10\ 12)$, the transpositions of the out-arcs of $r_1,r_2,c_1,c_2,s_1,s_2$. The default system has three closed trails. Switching at $r_2$, $c_2$ and $s_2$, that is, taking $S$ with $\tau_S=(3\ 4)(7\ 8)(10\ 12)$, gives $\tau_S\beta=(1\ 5\ 9)(2\ 8\ 10)(3\ 6\ 12)(4\ 7\ 11)$, the four triangles of the completion in \autoref{fig:latin}(a). Hence $\max_S c(\tau_S\beta)=4=|A(D_L)|/3$. The resulting instance of {\sc Swap Median} is
\[
\pi_1=\iota_{12},\quad \pi_2=\tau=[2\ 1\ 4\ 3\ 6\ 5\ 8\ 7\ 11\ 12\ 9\ 10],\quad \pi_3=\beta,
\]
with $k=6$ and $n=12$. Its median value is $\med(\Pi)=6+12-4=14$, attained by
\[\sigma=\tau_S=[1\ 2\ 4\ 3\ 5\ 6\ 8\ 7\ 9\ 12\ 11\ 10].\] In contrast, for the non-completable square with $1$ in cell $(1,1)$ and $2$ in cell $(2,2)$, $D_L$ is a single directed $6$-cycle, so the maximum is $1<6/3$.
\end{example}

\begin{theorem}\label{thm:swap-median-npc}
{\sc Swap Median} is \NP-complete for three input permutations. It remains \NP-complete when two of the input permutations differ by a set of disjoint swaps.
\end{theorem}

\begin{proof}
Membership in \NP is clear. Given a partial Latin square $L$, build $D'_L$ and apply \autoref{lem:transition} to it, obtaining $\beta$ and $k$ disjoint transpositions on $n=|A(D'_L)|$ points. Let $\Pi=(\iota,\tau,\beta)$ and $f=k+n-|A(D_L)|/3$. By \autoref{thm:median-cube}, $\med(\Pi)=k+n-\max_S c(\tau_S\beta)$. By \autoref{lem:transition} and Step~2, $\max_S c(\tau_S\beta)$ is the maximum number of cycles in a cycle decomposition of $D_L$. This number is at most $|A(D_L)|/3$, with equality if and only if $L$ is completable (\autoref{lem:latin}). Hence $\med(\Pi)\leq f$ if and only if $L$ is completable. The construction is polynomial.
\end{proof}

In the instances of \autoref{thm:swap-median-npc}, an optimal median lies on a geodesic between $\pi_1$ and $\pi_2$. Moreover, any optimal median can be turned in polynomial time into one of the form $\tau_S$, because the proof of \autoref{lem:cube-projection} is constructive. The hardness therefore comes entirely from the third input: choosing the best point of a Boolean cube with respect to $\pi_3$ is already hard.

%% file: figures/fig_latin.tex
\begin{tikzpicture}[scale=1.0, >={Stealth[length=2mm]}]
  \node[font=\small] at (-0.35,1.55) {(a)};
  \draw (0,0) grid (1,1) ; \draw (0,0) rectangle (1,1);
  \draw[step=0.5] (0,0) grid (1,1);
  \node[font=\scriptsize] at (0.5,-0.25) {$L$ (empty)};
  \begin{scope}[xshift=1.5cm]
    \draw[step=0.5] (0,0) grid (1,1);
    \node at (0.25,0.75) {$1$}; \node at (0.75,0.75) {$2$};
    \node at (0.25,0.25) {$2$}; \node at (0.75,0.25) {$1$};
    \node[font=\scriptsize] at (0.5,-0.25) {completion};
  \end{scope}
  \begin{scope}[xshift=4.0cm, yshift=-0.2cm]
    \node[font=\small] at (-0.6,1.75) {(b)};
    \node[gvert] (r1) at (0,1.3) {$r_1$}; \node[gvert] (r2) at (0,0) {$r_2$};
    \node[gvert] (c1) at (2,1.3) {$c_1$}; \node[gvert] (c2) at (2,0) {$c_2$};
    \node[gvert] (s1) at (4,1.3) {$s_1$}; \node[gvert] (s2) at (4,0) {$s_2$};
    \tikzset{lab/.style={font=\tiny, fill=white, inner sep=0.5pt}}
    \draw[->] (r1) -- node[lab, pos=0.5]{1} (c1);
    \draw[->] (r1) -- node[lab, pos=0.3]{2} (c2);
    \draw[->] (r2) -- node[lab, pos=0.3]{3} (c1);
    \draw[->] (r2) -- node[lab, pos=0.5]{4} (c2);
    \draw[->] (c1) -- node[lab, pos=0.5]{5} (s1);
    \draw[->] (c1) -- node[lab, pos=0.3]{6} (s2);
    \draw[->] (c2) -- node[lab, pos=0.3]{7} (s1);
    \draw[->] (c2) -- node[lab, pos=0.5]{8} (s2);
    \draw[->] (s1) to[out=120, in=60] node[lab, pos=0.5]{9} (r1);
    \draw[->] (s2) to[out=105, in=35] node[lab, pos=0.62]{10} (r1);
    \draw[->] (s1) to[out=-105, in=-35] node[lab, pos=0.62]{11} (r2);
    \draw[->] (s2) to[out=-120, in=-60] node[lab, pos=0.5]{12} (r2);
  \end{scope}
  \begin{scope}[xshift=9.6cm, yshift=0.05cm]
    \node[font=\small] at (-0.4,1.5) {(c)};
    \foreach \y/\l in {1.2/1,0.6/2,0/3} {
      \draw (0,\y) -- (2.4,\y);
      \node[font=\tiny, left] at (0,\y) {in$_\l$}; \node[font=\tiny, right] at (2.4,\y) {out$_\l$};
    }
    \foreach \x/\a/\b in {0.6/1.2/0.6, 1.2/0.6/0, 1.8/1.2/0.6} {
      \draw[line width=1pt] (\x,\a) -- (\x,\b); \fill (\x,\a) circle (1.6pt); \fill (\x,\b) circle (1.6pt);
    }
    \node[font=\scriptsize] at (1.2,-0.4) {$d=3$: three switches};
  \end{scope}
\end{tikzpicture}

%% file: sections/algorithms.tex
\section{Polynomial-time algorithms for short cycles}
\label{sec:algorithms}

The blue cycles in \autoref{thm:cgp-hard} have length at most four. We show that this is the threshold: if one of the two relative permutations has all its cycles of length at most three, a maximum common lower bound can be computed efficiently.

\begin{theorem}\label{thm:short-cycles}
Let $R,B\in S_n$ and assume that every cycle of $B$ has length at most $3$. Then a common lower bound of $R$ and $B$ of maximum rank can be computed in $O(n^2)$ time.
\end{theorem}

\begin{proof}
By \autoref{thm:biane}, a permutation $\lambda$ is a common lower bound of $R$ and $B$ if and only if its non-trivial cycles have element sets $X_1,\ldots,X_p$ satisfying the following conditions.
(a) Each $X_j$ is contained in a cycle $C$ of $R$ and in a cycle $D$ of $B$.
(b) The sets contained in a common red cycle are pairwise non-crossing with respect to it, and the same holds for blue cycles.
(c) Each $X_j$ carries the same cyclic order in $C$ and in $D$, and the cycle of $\lambda$ on $X_j$ is this order.
The rank of $\lambda$ is $\sum_j(|X_j|-1)$.

Since $|D|\leq 3$, two disjoint sets of size at least two cannot lie in the same blue cycle, and crossing requires four elements. Hence the blue part of (b) is automatic. Call a set $X$ with $|X|\in\{2,3\}$ a \emph{candidate} if it is contained in a red cycle and in a blue cycle and, when $|X|=3$ (and thus $X=D$), the cyclic orders induced on $X$ by $R$ and by $B$ coincide. Condition (c) is automatic for $|X|=2$. Every element belongs to at most three candidates, namely the subsets of size at least two of its blue cycle that contain it. Hence there are $O(n)$ candidates, and they can be listed in $O(n)$ time.

The problem therefore splits into independent problems, one per red cycle. Fix such a cycle $C=(c_0\ c_1\ \cdots\ c_{L-1})$. We must choose pairwise disjoint and non-crossing candidates contained in $C$, maximizing the total weight $\sum(|X|-1)$. Cutting the cycle between $c_{L-1}$ and $c_0$ does not change which sets cross, because crossing is defined by alternation. For $0\leq i\leq j<L$, let $f(i,j)$ be the optimum using only candidates contained in $\{c_i,\ldots,c_j\}$, with $f(i,j)=0$ when $i>j$. If $c_i$ is not covered, the value is $f(i+1,j)$. Otherwise $c_i$ is the first element of a candidate $X$ with positions $i=p_0<p_1<\cdots<p_q\leq j$. Non-crossing then means that every other chosen candidate lies inside one of the gaps $(p_{a},p_{a+1})$ or after $p_q$. Hence
\[
f(i,j)=\max\Big\{f(i+1,j),\ \max_{X}\Big(|X|-1+\sum_{a=0}^{q-1}f(p_a+1,p_{a+1}-1)+f(p_q+1,j)\Big)\Big\},
\]
where $X$ ranges over the at most three candidates whose first position is $i$ and whose last position is at most $j$. The table has $O(L^2)$ entries, each computed in constant time, and a standard traceback recovers an optimal family. Summing over red cycles gives $O(n^2)$ time.
\end{proof}

\autoref{fig:dp} illustrates the algorithm. Let $R=(1\ 2\ 3\ 4\ 5\ 6)$ and $B=(1\ 3\ 5)(2\ 4)(6)$. The candidates are the pairs $\{1,3\}$, $\{1,5\}$, $\{3,5\}$, $\{2,4\}$ and the triple $\{1,3,5\}$, whose cyclic order $(1\ 3\ 5)$ is the same in $R$ and in $B$. The triple has weight $2$ but crosses $\{2,4\}$ in the red cycle, and the nested pairs $\{1,5\}$ and $\{2,4\}$ also have total weight $2$. No family has weight $3$, so the maximum rank is $2$. For $\Pi=(\iota,R,B)$ we have $\dd_{12}=5$, $\dd_{13}=3$ and $\dd_{23}=4$, hence $\beta_1=2$, and $\sigma=(1\ 5)(2\ 4)=[5\ 4\ 3\ 2\ 1\ 6]$ attains $\LB=6$.

\begin{figure}[!h]
    \centering
    \input{figures/fig_dp}
    \caption{(a)~An instance with blue cycles of length at most $3$. Inside the red cycle $R$, the shaded triangle is the triple candidate $\{1,3,5\}$, and the dashed segments are the pair candidates. An optimal family, $\{1,5\}$ and $\{2,4\}$, is shown in green. (b)~The recurrence of \autoref{thm:short-cycles} on a cut red cycle. When $c_i$ is the first element of a chosen candidate $X$, every other chosen candidate lies in one of the gaps of $X$, and the gaps are solved independently.}
    \label{fig:dp}
\end{figure}

By symmetry, the same holds if $R$ has all cycles of length at most $3$. Together with \autoref{thm:cgp-hard}, this gives a dichotomy.

\begin{corollary}[dichotomy]\label{cor:dichotomy}
{\sc Common Geodesic Prefix} is solvable in polynomial time when one of $\pi_1^{-1}\pi_2$ and $\pi_1^{-1}\pi_3$ has all cycles of length at most $3$, and it is \NP-complete (and \APX-hard in its maximization version) when the cycles of one of them have length at most $4$ and the cycles of the other have length at most $8$.
\end{corollary}

Recall from \autoref{thm:lb-characterization} that the attainment of the median lower bound can be tested at \emph{any} pivot. Every relative permutation $\pi_i^{-1}\pi_j$ appears at the pivot $i$, and its inverse, which has the same cycle lengths, appears at the pivot $j$.

\begin{corollary}\label{cor:lb-poly}
If some relative permutation $\pi_i^{-1}\pi_j$ ($i\neq j$) has all cycles of length at most $3$, one can decide in $O(n^2)$ time whether $\med(\Pi)=\LB(\Pi)$, and if so compute a median. Moreover, the pivot values $P_i$ and $P_j$ can be computed in $O(n^2)$ time.
\end{corollary}

In particular, \autoref{cor:lb-poly} applies when two of the input genomes differ by disjoint swaps or by short cyclic exchanges, which is a natural situation when two genomes are closely related.

%% file: figures/fig_dp.tex
\begin{tikzpicture}[
    pt/.style={circle, fill=black, inner sep=0pt, minimum size=3.5pt},
    lab/.style={font=\scriptsize, inner sep=1pt}]
  \node[font=\small] at (-1.35,1.05) {(a)};
  \drawcycle{R}{0}{0}{0.85}{1,2,3,4,5,6}{red!75!black}
  \node[font=\scriptsize, text=red!75!black] at (-1.15,-0.85) {$R$};
  \fill[blue!12] (R-1)--(R-3)--(R-5)--cycle;
  \draw[blue!70!black, line width=0.6pt, dashed] (R-1)--(R-3)--(R-5)--cycle;
  \draw[blue!70!black, line width=0.6pt, dashed] (R-2)--(R-4);
  \draw[chordsel] (R-1)--(R-5);
  \draw[chordsel] (R-2)--(R-4);
  \foreach \x in {1,...,6} \fill[red!75!black] (R-\x) circle (1.3pt);
  \drawcycle{Ba}{2.15}{0.35}{0.42}{1,3,5}{blue!70!black}
  \drawcycle{Bb}{3.25}{0.35}{0.32}{2,4}{blue!70!black}
  \fill[blue!70!black] (3.95,0.35) circle (1.3pt);
  \node[lab] at (3.95,0.6) {$6$};
  \node[font=\scriptsize, text=blue!70!black] at (3.0,-0.55) {$B=(1\,3\,5)(2\,4)(6)$};
  \begin{scope}[xshift=4.85cm, yshift=-0.35cm]
    \node[font=\small] at (-0.15,1.4) {(b)};
    \draw[line width=0.7pt, black!50] (0,0) -- (7.2,0);
    \foreach \x/\l in {0.3/c_i, 2.4/c_{p_1}, 4.5/c_{p_2}, 6.9/c_j} {
      \node[pt] at (\x,0) {};
      \node[lab, below=2pt] at (\x,0) {$\l$};
    }
    \foreach \x in {0.95,1.5,1.95, 3.0,3.55,4.0, 5.1,5.65,6.25} \node[pt, fill=black!45] at (\x,0) {};
    \draw[chordsel] (0.3,0) to[out=60, in=120] (2.4,0);
    \draw[chordsel] (2.4,0) to[out=60, in=120] (4.5,0);
    \node[font=\scriptsize, text=black!60!green] at (2.4,1.0) {$X$, weight $|X|-1$};
    \foreach \a/\b/\t in {0.8/2.1/{f(i{+}1,p_1{-}1)}, 2.9/4.2/{f(p_1{+}1,p_2{-}1)}, 4.9/7.2/{f(p_2{+}1,j)}} {
      \draw[decorate, decoration={brace, amplitude=4pt, mirror}, line width=0.6pt] (\a,-0.45) -- (\b,-0.45);
      \node[font=\scriptsize] at ({(\a+\b)/2},-0.85) {$\t$};
    }
    \draw[line width=0.8pt, black!60] (0.95,0) to[out=70, in=110] (1.95,0);
    \draw[line width=0.8pt, black!60] (5.1,0) to[out=70, in=110] (6.25,0);
  \end{scope}
\end{tikzpicture}

%% file: sections/approximation.tex
\section{Approximation and inapproximability}
\label{sec:approximation}

Returning the best input gives a $4/3$-approximation for {\sc Swap Median} with three inputs and a $2$-approximation for {\sc Swap Closest}, since $\min_r D_r\leq \frac43\LB$ and $\min_r\max\{\dd_{rs},\dd_{rt}\}\leq Q\leq 2\,\LB_{\clo}$; both bounds hold in every metric space~\cite{bachmaier2015hardness}. The pivot framework of \autoref{sec:geodesic} refines this in an \emph{a posteriori} way.

\begin{proposition}[a posteriori certificate]\label{prop:certificate}
Let $\lambda$ be any common lower bound at pivot $r$, computed by any algorithm, and let $\sigma=\pi_r\lambda$. Then
\[
F_\Pi(\sigma)=\LB(\Pi)+\big(\beta_r-\ellp(\lambda)\big)\leq \med(\Pi)+\big(\beta_r-\ellp(\lambda)\big).
\]
In particular, $\sigma$ is optimal when $\ellp(\lambda)=\beta_r$, and if $r$ minimizes $D_r$ then $F_\Pi(\sigma)/\med(\Pi)\leq \frac43-\ellp(\lambda)/\LB(\Pi)$. If moreover $\ellp(\lambda)=M_r$ for every $r$ and $\med(\Pi)\leq\LB(\Pi)+1$, the best of the three pivot solutions is optimal.
\end{proposition}

\begin{proof}
The equality is \autoref{lem:pivot-value} together with $D_r-\beta_r=\LB$, and the inequality uses $\LB\leq\med$. If $D_r\leq\frac43\LB$, then $F_\Pi(\sigma)=D_r-\ellp(\lambda)\leq\frac43\LB-\ellp(\lambda)$, and we divide by $\med\geq\LB$. The last claim is \autoref{thm:small-excess}.
\end{proof}

Common lower bounds can be produced by any independent set of the crossing graph (\autoref{lem:is-to-clb}), by the exact algorithm of \autoref{thm:short-cycles} when it applies, or by the greedy geodesic walk of \autoref{sec:experiments}. For {\sc Swap Closest}, applying $h\leq\min\{\ellp(\lambda),\lfloor C_r/2\rfloor\}$ swaps of $\lambda$ from the input $\pi_r$ minimizing $C_r=\max\{\dd_{rs},\dd_{rt}\}$ gives value $C_r-h$, hence ratio at most $2-2h/Q$. These guarantees depend on the instance. \autoref{thm:cgp-hard} shows that the underlying optimization problem, maximizing $\ellp(\lambda)$, is \APX-hard. So this route cannot give a polynomial-time approximation scheme for the pivot value, unless $\P=\NP$.

\paragraph*{A robustness conjecture.}
Uniform guarantees and inapproximability results would follow from a quantitative version of \autoref{thm:small-excess}. That theorem says that a solution with total defect at most $2$ is a pivot solution. The conjecture below says that a solution with small defects with respect to two pairs is close to a pivot solution. For permutations $\sigma,\tau$, let $2\,\gamma(\sigma,\tau)=\ellp(\sigma)+\ellp(\sigma^{-1}\tau)-\ellp(\tau)\geq 0$, so that $\gamma(\sigma,\tau)=0$ if and only if $\sigma\lea\tau$.

\begin{conjecture}\label{conj:robust}
For all $\sigma,\tau_2,\tau_3\in S_n$ there is a permutation $\lambda$ with $\lambda\lea\sigma$, $\lambda\lea\tau_2$ and $\lambda\lea\tau_3$ such that
\[
\ellp(\lambda)\ \geq\ \ellp(\sigma)-3\big(\gamma(\sigma,\tau_2)+\gamma(\sigma,\tau_3)\big).
\]
\end{conjecture}

We know no counterexample to \autoref{conj:robust}, and the constant $3$ cannot be lowered: for $\sigma=(1\,4)(2\,6)(3\,5)$, $\tau_2=(1\,5\,6)(2\,3\,4)$ and any $\tau_3\succeq\sigma$, we have $\gamma(\sigma,\tau_2)=1$ while the only common lower bound of $\sigma$ and $\tau_2$ is $\iota$. The conjecture would have the following consequences.

\begin{theorem}\label{thm:conditional}
If \autoref{conj:robust} holds, then:
\begin{enumerate}
\item for every instance, $\min_r P_r\leq \med(\Pi)+\frac23\big(\med(\Pi)-\LB(\Pi)\big)$, so the best exact pivot solution is a $\frac{10}{9}$-approximation for {\sc Swap Median};
\item {\sc Swap Closest} with three input permutations is \APX-hard.
\end{enumerate}
\end{theorem}

\begin{proof}
(1) Let $\sigma$ be an optimal median with defects $e_{ij}$, and let $s=\med-\LB=(e_{12}+e_{13}+e_{23})/2$. For a pivot $r$, apply the conjecture to $\lambda_0=\pi_r^{-1}\sigma$, $\tau^r_s$ and $\tau^r_t$. By left-invariance, $2\gamma(\lambda_0,\tau^r_s)=e_{rs}$ and $2\gamma(\lambda_0,\tau^r_t)=e_{rt}$. Moreover $\ellp(\lambda_0)=\dd(\pi_r,\sigma)=\beta_r+(e_{rs}+e_{rt}-e_{st})/2$. Hence $M_r\geq\beta_r-(e_{rs}+e_{rt})-e_{st}/2$, that is, $g_r\leq e_{rs}+e_{rt}+e_{st}/2$. Summing over $r$, each defect appears with total coefficient $5/2$, so $\sum_r g_r\leq 5s$ and $\min_r g_r\leq\frac53 s$. Thus $\min_rP_r=\LB+\min_r g_r\leq\med+\frac23(\med-\LB)$. Writing $x=\LB/\med\in[\frac34,1]$, we get $\min_r P_r/\med\leq\min\{1+\frac23(1-x),\frac43x\}\leq\frac{10}{9}$, with equality at $x=\frac56$.

(2) We use the gap version of the \APX-hardness of {\sc Maximum Independent Set} on cubic graphs~\cite{alimonti2000some,berman1999some}: there is $\varepsilon_0>0$ such that, given a cubic graph $G$ on $n$ vertices and an integer $a\geq n/4$, it is \NP-hard to distinguish $\alpha(G)\geq a$ from $\alpha(G)<(1-\varepsilon_0)a$. Let $m=3n/2$, $\Pi=\Pi_G$ (\autoref{constr:subdivision}), $k=a+m$, and let $(\Pi'_k,f)$ be given by \autoref{constr:closest}. If $\alpha(G)\geq a$, then $\clo(\Pi'_k)=f$ by \autoref{thm:closest-equivalence}. Conversely, let $\sigma'$ satisfy $\dd(\sigma',\pi'_i)\leq f+\delta$ for all $i$. Then $e_{12}(\sigma')\leq 2\delta$, $e_{13}(\sigma')\leq 2\delta$ and $\dd(\pi'_1,\sigma')\geq 2f-(f+\delta)=f-\delta$. The conjecture gives a common lower bound of $\pi_1'^{-1}\pi'_2$ and $\pi_1'^{-1}\pi'_3$ of rank at least $f-\delta-6\delta$, and the block argument of \autoref{thm:closest-equivalence} turns it into a common lower bound in block $A$ of rank at least $k-7\delta$. Hence $M_1(\Pi_G)=\alpha(G)+m\geq k-7\delta$. If $\alpha(G)<(1-\varepsilon_0)a$, we get $\delta>\varepsilon_0a/7\geq\varepsilon_0 n/28$. Since $f=x+k\leq \dd_{12}+k\leq (n+4m)+(n+m)=9.5\,n$, the ratio $\clo(\Pi'_k)/f$ exceeds $1+\varepsilon_0/266$ in the no case, which rules out a PTAS unless $\P=\NP$.
\end{proof}

We emphasize that \autoref{thm:conditional} is conditional. Proving \autoref{conj:robust}, even with a larger constant, would settle the approximability of {\sc Swap Closest} for three inputs. Whether {\sc Swap Median} for three inputs is \APX-hard is also open: the reduction of \autoref{sec:median} is not approximation preserving, because the optimum value $k+n-\max_S c(\tau_S\beta)$ is dominated by the term $k+n$.

%% file: sections/experiments.tex
\section{Experiments on simulated data}
\label{sec:experiments}

We implemented the methods of this paper in Python. We also wrote an exhaustive C solver, which computes $\med$, $\clo$ and all pivot values for $n\leq 12$. The code is available in the repository accompanying the paper.

\paragraph*{Algorithms.}
\emph{Best input} returns the input that minimizes the objective. The \emph{pivot heuristic} runs, from each pivot $r$, a \emph{greedy geodesic walk}. Starting from $\lambda=\iota$, it repeatedly applies a swap $(a\ b)$ such that $a,b$ lie in a common cycle of $\lambda^{-1}\tau^r_s$ and in a common cycle of $\lambda^{-1}\tau^r_t$. Such a swap splits a cycle of both (\autoref{lem:swap-effect}), so $\lambda$ remains a common lower bound. Among the available swaps, the walk prefers one whose split is most unbalanced in both cycles, because short chords cross few others. The walk stops at a maximal common lower bound. The best of the three resulting pivot solutions is returned, together with its certificate $\beta_r-\ellp(\lambda)$ (\autoref{prop:certificate}). \emph{Local search} starts from the pivot solution and from the inputs and performs iterated descent over single swaps. A swap of positions $x,y$ decreases $\dd(\sigma,\pi_i)$ if and only if $x,y$ lie in a common cycle of $\sigma^{-1}\pi_i$, so improving swaps for the median are found in linear time by bucketing elements according to their cycles in two of the relative permutations. For {\sc Swap Closest}, local search minimizes the maximum distance, breaking ties by the sum.

\paragraph*{Data.}
In the \emph{uniform} model the three inputs are independent uniformly random permutations. In the \emph{evolutionary} model with rate $\rho$, a random ancestor permutation is drawn, and each of the three inputs is obtained from it by $\max\{1,\mathrm{round}(\rho n)\}$ random swaps (a star phylogeny). In this model the ancestor is a natural reference solution.

\paragraph*{Small instances: comparison with the optimum.}
\autoref{tab:e1} reports $1\,560$ instances with $n\in\{6,8,10\}$ solved exactly. Every instance satisfies $\med\leq\LB+1$. By \autoref{thm:small-excess}, the optimum is therefore always a pivot solution, and the greedy pivot heuristic actually found an optimal median on all instances. The median lower bound is attained in $77.5\%$ to $100\%$ of the instances, less often for larger $n$ and for more divergent inputs. Returning the best input loses $18$--$29\%$ on average, close to the worst-case factor $4/3$. For {\sc Swap Closest}, the lower bound $\lceil Q/2\rceil$ is attained in at least $95\%$ of the instances, local search is always optimal, and the best input is often close to the worst-case factor $2$.

\begin{table}[!h]
\centering
\setlength{\tabcolsep}{2.4pt}
\footnotesize
\begin{tabular}{rlr|rrrrr|rrr}
\toprule
& & & \multicolumn{5}{c|}{\sc Swap Median} & \multicolumn{3}{c}{\sc Swap Closest}\\
$n$ & model & inst. & $\med{=}\LB$ & $\med{\leq}\LB{+}1$ & pivot opt. & LS opt. & input/opt & $\clo{=}\LB_{\clo}$ & LS opt. & input/opt\\
\midrule
6 & uniform & 200 & 96.5 & 100.0 & 100.0 & 100.0 & 1.182 & 97.0 & 100.0 & 1.522 \\
6 & evo $0.15$ & 200 & 100.0 & 100.0 & 100.0 & 100.0 & 1.267 & 100.0 & 100.0 & 1.985 \\
6 & evo $0.3$ & 200 & 100.0 & 100.0 & 100.0 & 100.0 & 1.178 & 100.0 & 100.0 & 1.730 \\
6 & evo $0.6$ & 200 & 95.5 & 100.0 & 100.0 & 100.0 & 1.227 & 97.5 & 100.0 & 1.892 \\
8 & uniform & 150 & 91.3 & 100.0 & 100.0 & 100.0 & 1.213 & 98.0 & 100.0 & 1.646 \\
8 & evo $0.15$ & 150 & 100.0 & 100.0 & 100.0 & 100.0 & 1.293 & 100.0 & 100.0 & 2.000 \\
8 & evo $0.3$ & 150 & 100.0 & 100.0 & 100.0 & 100.0 & 1.233 & 100.0 & 100.0 & 1.853 \\
8 & evo $0.6$ & 150 & 88.7 & 100.0 & 100.0 & 100.0 & 1.212 & 98.7 & 100.0 & 1.789 \\
10 & uniform & 40 & 77.5 & 100.0 & 100.0 & 100.0 & 1.226 & 97.5 & 100.0 & 1.711 \\
10 & evo $0.15$ & 40 & 100.0 & 100.0 & 100.0 & 100.0 & 1.292 & 100.0 & 100.0 & 1.975 \\
10 & evo $0.3$ & 40 & 97.5 & 100.0 & 100.0 & 100.0 & 1.225 & 100.0 & 100.0 & 1.833 \\
10 & evo $0.6$ & 40 & 77.5 & 100.0 & 100.0 & 100.0 & 1.229 & 95.0 & 100.0 & 1.783 \\
\bottomrule
\end{tabular}
\caption{Exact comparison on small instances (percentages of instances, and average ratios for the best input). ``pivot opt.'' and ``LS opt.'' give the percentage of instances on which the greedy pivot heuristic and local search return an optimal solution.}
\label{tab:e1}
\end{table}

\begin{figure}[!h]
    \centering
    \includegraphics[width=\textwidth]{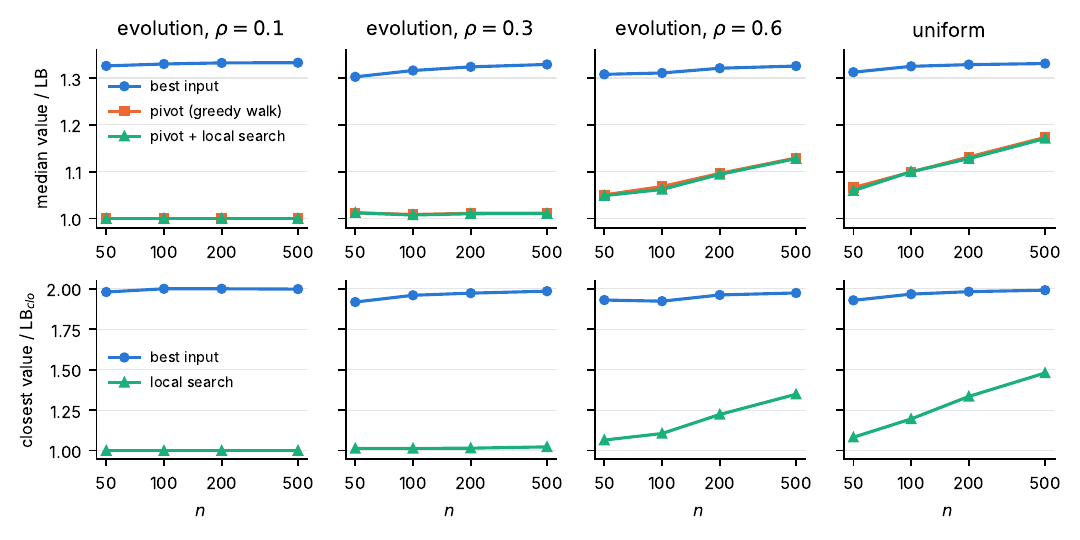}
    \caption{Larger instances: average ratio between the value of the computed solution and the lower bound ($\LB$ for the median, top; $\LB_{\clo}$ for the closest problem, bottom), for the evolutionary model with rates $\rho\in\{0.1,0.3,0.6\}$ and for uniform inputs.}
    \label{fig:e2}
\end{figure}

\paragraph*{Larger instances.}
For $n\in\{50,100,200,500\}$ ($20$ instances per point) the optimum is unknown, and we report the ratio between the value found and the lower bound (\autoref{fig:e2}). This ratio is therefore an \emph{upper} bound on the true approximation ratio. For low and moderate divergence ($\rho\in\{0.1,0.3\}$), the pivot heuristic is within $1.5\%$ of $\LB$, and it attains $\LB$ (hence is provably optimal) on all instances with $\rho=0.1$. In this regime the median also recovers the ancestor: for $n=500$ and $\rho=0.3$, the average distance between the computed median and the true ancestor is $13.8$, compared with $149.2$ for the closest input; for $\rho=0.1$ it is $0.05$, compared with $49.9$. For high divergence and for uniform inputs, the gap to $\LB$ grows with $n$. Since $\LB$ is likely weak there, these values do not determine how far the heuristics are from optimal. Local search improves the pivot solution only marginally on the median objective, but substantially on the closest objective, where it reduces the ratio from about $2$ (best input) to between $1.00$ and $1.48$. Local search for the median takes less than $4$ seconds per instance for $n=500$.

These experiments support the practical relevance of the geodesic viewpoint. On biologically plausible instances, where the inputs share a recent common ancestor, pivot solutions certify optimality through $\LB$ and recover the ancestor. The hard instances behind \autoref{thm:cgp-hard}, \autoref{thm:swap-closest-npc} and \autoref{thm:swap-median-npc} are structured, and they do not arise in these random models.

%% file: sections/conclusion.tex
\section{Conclusion and open problems}
\label{sec:conclusion}

We proved that {\sc Swap Median} and {\sc Swap Closest} are \NP-complete for three input permutations, the smallest non-trivial case.

For the median, the key ingredient is a projection lemma. When two inputs differ by disjoint swaps, some optimal median lies on a geodesic between them. The problem then becomes a maximum cycle decomposition of an Eulerian digraph, and Colbourn's theorem on partial Latin squares gives the hardness.

For the closest problem, the hardness comes from the absolute order of the symmetric group. Median candidates on two geodesics leaving a common input are common lower bounds in this order, and finding the deepest one is \NP-hard and \APX-hard. Padding transfers this hardness to {\sc Swap Closest}, even for deciding attainment of the trivial lower bound.

We also characterized when the triangle lower bound of the median is attained, and we showed that every solution exceeding it by at most one is a pivot solution, whereas optimal medians are not pivot solutions in general. Finally, we identified polynomial cases through a dichotomy on cycle lengths, and we obtained a simple certificate-based heuristic that is optimal on most simulated instances. Several questions remain open.

\begin{enumerate}
\item \emph{Lower-bound attainment for the median.} What is the complexity of deciding whether $\med(\Pi)=\LB(\Pi)$, that is, whether $M_1=\beta_1$? By \autoref{cor:lb-poly}, it is polynomial when one relative permutation has short cycles. The instances of \autoref{sec:median} need not attain the lower bound, so they do not answer this question.

\item \emph{Approximability.} Is {\sc Swap Median} (or {\sc Swap Closest}) with three inputs \APX-hard? Proving \autoref{conj:robust} would settle the closest case. For the median, an approximation-preserving variant of \autoref{sec:median} would be needed. Conversely, is there a polynomial-time approximation better than $4/3$ for the median?

\item \emph{Geodesic medians.} Does every instance have an optimal median on a geodesic between two of the inputs? This holds in the instances of \autoref{thm:swap-median-npc}. However, \autoref{prop:pivot-not-optimal} shows that one cannot require two geodesics leaving a common input.

\item \emph{Parameterized complexity.} Is {\sc Common Geodesic Prefix} fixed-parameter tractable with respect to $k$, or with respect to the gap $\beta_1-k$?
\end{enumerate}

Finally, we compare with Kendall's tau distance. Kemeny aggregation was recently shown to be \NP-hard for three rankings~\cite{madarasi2026kemeny,peters2026kemeny}, through a reduction from max-cut in which every pair of candidates is split two-to-one. For Kendall's tau, the analogue of our pivot problem is easy, because geodesics correspond to intervals of the weak order, which is a lattice~\cite{bjorner2005combinatorics}. Our two proofs exploit features specific to the swap distance. The absolute order is not a lattice, which gives the hardness of the pivot problem and of {\sc Swap Closest}. And cycle counts of products of a permutation with disjoint transpositions encode transition systems of Eulerian digraphs, which gives the hardness of {\sc Swap Median}.

%% file: sections/appendix.tex
\section{The instances of \autoref{prop:pivot-not-optimal}}
\label{app:examples}

Permutations are given in one-line notation. The pivot values $P_r=D_r-M_r$ are obtained from \autoref{thm:biane}, by listing the common lower bounds of the two relative permutations at each pivot.

\paragraph*{Two blocks.}
\[
\begin{array}{rcl}
\pi_1&=&[4\ 1\ 5\ 2\ 6\ 3\ 11\ 12\ 10\ 9\ 7\ 8],\\
\pi_2&=&[5\ 6\ 4\ 3\ 1\ 2\ 10\ 7\ 11\ 8\ 12\ 9],\\
\pi_3&=&[3\ 5\ 1\ 6\ 2\ 4\ 9\ 11\ 7\ 12\ 8\ 10].
\end{array}
\]
The instance is the disjoint union of a $6$-element instance whose unique best pivot is $\pi_1$ and of the same instance with the roles of $\pi_1$ and $\pi_2$ exchanged. The pairwise distances are $\dd_{12}=6$, $\dd_{13}=7$ and $\dd_{23}=7$, so $\LB=10$ and $(\beta_1,\beta_2,\beta_3)=(3,3,4)$. The pivot values are $(P_1,P_2,P_3)=(13,13,14)$. The permutation
\[
\sigma=[4\ 1\ 5\ 2\ 6\ 3\ 10\ 7\ 11\ 8\ 12\ 9]
\]
satisfies $\dd(\sigma,\pi_1)=3$, $\dd(\sigma,\pi_2)=3$ and $\dd(\sigma,\pi_3)=6$, so $F_\Pi(\sigma)=12<\min_r P_r$.

\paragraph*{A connected instance.}
Exchanging the entries at positions $1$ and $7$ of $\pi_2$ above merges the two blocks:
\[
\begin{array}{rcl}
\pi_1&=&[4\ 1\ 5\ 2\ 6\ 3\ 11\ 12\ 10\ 9\ 7\ 8],\\
\pi_2&=&[10\ 6\ 4\ 3\ 1\ 2\ 5\ 7\ 11\ 8\ 12\ 9],\\
\pi_3&=&[3\ 5\ 1\ 6\ 2\ 4\ 9\ 11\ 7\ 12\ 8\ 10].
\end{array}
\]
Here $\dd_{12}=7$, $\dd_{13}=7$ and $\dd_{23}=8$, so $\LB=11$ and $(\beta_1,\beta_2,\beta_3)=(3,4,4)$. The group generated by $\pi_1^{-1}\pi_2$ and $\pi_1^{-1}\pi_3$ is transitive, and $(P_1,P_2,P_3)=(14,14,15)$. The same permutation $\sigma$ as above satisfies $\dd(\sigma,\pi_1)=3$, $\dd(\sigma,\pi_2)=4$ and $\dd(\sigma,\pi_3)=6$, so $F_\Pi(\sigma)=13<\min_r P_r$.